\documentclass{gGAF2e}       

\usepackage{amsmath}
\usepackage{graphicx}
\usepackage{subfigure}

\usepackage{mathptmx}      
\usepackage{captcont}

\usepackage[nomarkers, nolists,figuresonly]{endfloat}
\usepackage[usenames,dvipsnames,svgnames,table]{xcolor}

\newsavebox{\astrutbox}

\sbox{\astrutbox}{\rule[-5pt]{0pt}{20pt}}

\def\der#1#2{{\partial #1\over \partial #2}}

\newcommand{\be}{\begin{equation}}
\newcommand{\ee}{\end{equation}}

\def\bea{\begin{eqnarray}}

\def\eea{\end{eqnarray}}

\def\bse{\begin{subequations}}

\def\ese{\end{subequations}}

\def\bsea{\begin{subeqnarray}}

\def\esea{\end{subeqnarray}}

\def\({\left (}

\def\){\right )}

\def\[{\left [}

\def\]{\right ]}

\def\<{\left <}

\def\>{\right >}

\newcommand{\beq}{\begin{equation}}
\newcommand{\eeq}{\end{equation}}
\def\la{\hbox{\raise.35ex\rlap{$<$}\lower.6ex\hbox{$\sim$}\ }}
\def\ga{\hbox{\raise.35ex\rlap{$>$}\lower.6ex\hbox{$\sim$}\ }}

\def\beq{\begin{equation}}
\def\eeq{\end{equation}}
\def\beqa{\begin{eqnarray}}
\def\eeqa{\end{eqnarray}}
\def\sub#1{_{_{#1}}}

\def\be{\begin{equation}}
\def\eeq{\end{equation}}
\def\ba{\begin{align}}
\def\ea{\end{align}}

\jvol{00} \jnum{00} \jyear{2012} 

\title{On the mechanism of self gravitating Rossby interfacial waves in proto-stellar accretion discs}

\author{Ron Yellin-Bergovoy ${\dag}$$^{1}$\thanks{$\dag$Corresponding author. Email: yellinr@post.tau.ac.il
\vspace{6pt}} , Eyal Heifetz$^{1}$ \& Orkan M. Umurhan${^2}$\\\vspace{6pt}  ${^1}$Department of Earth Sciences, Tel-Aviv University, Tel-Aviv, Israel\\ ${^2}$NASA Ames Research Center, Division of Space Sciences, Planetary Systems Branch, Moffett Field, CA, 94035 USA\\\vspace{6pt}
\received{v1.0 submitted March 2015 } }

\begin{document}
\maketitle

\begin{abstract}

The dynamical response of edge waves under the influence of self-gravity is examined
in an idealized two-dimensional model of a proto-stellar disc, characterized
in steady state as a
rotating vertically infinite cylinder of fluid with constant density except
for a single density interface at some radius $r_0$.  The
fluid in basic state is prescribed to rotate with a Keplerian profile $\Omega_k(r)\sim r^{-3/2}$ modified by some additional azimuthal sheared flow. 
A linear analysis shows that there are two azimuthally propagating
edge waves, kin to the familiar Rossby waves and surface gravity waves in terrestrial studies, which
move opposite to one another with respect to the local
basic state rotation rate at the interface. Instability only occurs if the radial pressure gradient is opposite to that
of the density jump (unstably stratified) where self-gravity
acts as a wave stabilizer irrespective of the stratification of the system. The propagation properties
of the waves are discussed in detail
in the language of vorticity edge waves. The roles of both Boussinesq and non-Boussinesq effects
upon the stability and propagation of these waves with and without the inclusion
of self-gravity are then quantified. 
The dynamics involved with self-gravity non-Boussinesq effect is shown to be a source of vorticity production where there is a jump in the basic state density In addition, self-gravity also alters the dynamics via the radial main pressure gradient, which is  a Boussinesq effect .
Further applications of these mechanical insights are presented in
the conclusion including the ways in which multiple density jumps or gaps may or may not be stable.




\end{abstract}

\section{Introduction}

\label{intro}
The Atacama Large Millimeter/submillimeter Array (ALMA) has opened up a new
window revealing the
structure of proto-planet accretion discs (pp-discs hereafter). Most notable
of the recently reported ALMA discoveries 
is the appearance of asymmetric ``peanut" features in the outer ($> 80$AU) regions of several
pp-disc systems including SAO 206462, SR 21, LKH$\alpha$ 330 OPH IRS 48
and HD 142527 \citep{casassus2013flows,fukagawa2013local,isella2013azimuthal,van2013major,perez2014large}.  The ALMA website recently published a
high resolution image of HL Tauri,
a debris-disc phase T-Tauri system, showing a 
disc with alternating axisymmetric rings and gaps with unambiguous non-axisymmetric features
set atop.  While planets have
yet to be detected in the HL Tauri system, the gap-ring features appear to indicate an ongoing process
similar to the dynamics responsible for the Kirkwood gaps in the Jupiter-Asteroid-belt system, 
in which an embedded (as yet unseen) planet drives
gap creation through resonant gravitational forcing of an otherwise structurally uniform Keplerian disc 
(A. Parker, private communication).
Furthermore, the furthest parts of these same discs are likely to be sufficiently cold enough that
the densities reached in their local disc midplane's are correspondingly high enough such that
self-gravitational effects should not be ignored. The dynamical stability of sheared dynamical structures like those
seen in HL Tauri is the main inspiration for this work: we seek to 
develop a theoretical framework that could be helpful in interpreting the dynamical nature of structures
emerging in mainly cold pp-disc systems in which self-gravity of the disc gas plays an important dynamical role.
\par
Some pp-disc systems exhibiting peanut features
also show indications of severe dust depletion within their inner radial locations.  It has been
suggested that pressure/density enhancements in the outer parts of the disc 
act as particle traps preventing the inner parts of these discs from having their
dust replenished by accretion from the outer parts.  
Interestingly, axisymmetric pressure enhancements, provided they meet minimum requirements upon their
amplitudes and radial girth, are known to undergo a barotropic
shear transition known as the Rossby Wave Instability 
(RWI hereafter, \citep{lovelace1999rossby,li2000rossby,li2001rossby}).  This instability
leads to the creation of coherent anticylonic 
vortex structures with wide azimuthal extent. Recent studies have shown that this instability can develop even as early as in the infall stage of the disk \citep{bae2015protoplanetary}.
From both theoretical and numerical/computational considerations it is also understood that
anticylonic vortex structures in pp-discs readily entrain dust through 
well-known drag-momentum exchange mechanisms \citep{barge1995did}.
While the origins of the density enhancement and the causal dilemma posed by the above scenario  
remains unsettled (see recent discussion
in \citet{flock2014gaps}), the fact remains that rings/gaps and (generally 
speaking) {\emph{nearly axisymmetric annular structures}} are the basic states upon which
dynamical evolution proceeds in pp-discs.
A detailed examination of the dynamical state of such structures from a physically mechanistic
point of view is therefore of interest to theoreticians and observers alike.
  While the dynamical stability of self-gravitating disc systems is a vast subject 
  that we cannot possibly cover here
(see the reviews of \citet{armitage2011dynamics}),  we are interested in examining the 
mechanistic nature of stability in self-gravitating zonal flows in accretion discs.
Steady axisymmetric disc zonal flows arise from the balance between the 
Coriolis and centrifugal effects and the radial dependencies in the pressure and self gravity. Indeed, inspection of
the HL Tauri system suggests that the flow within the rings and gaps should
exhibit pronounced zonal flow structure
since the distribution of matter appears nearly axisymmetric, yet radially non-uniform.\par
Barotropic  zonal flows \footnote[1]{The term barotropic in this context refers to lack of variation of the basic shear flow in the vertical direction (i.e, perpendicular to the disk mid-plane). The RWI operates whether or not the disturbances are barotropic or baroclinic, but this is different from whether or not the basic flow is fully barotropic, fully baroclinic or pseudo-baroclinic.} can become unstable and the RWI is one example.  Barotropic shear instabilities,
i.e., instabilities in the absence of baroclinic torque but in the presence of significance rotation
may be interpreted within the framework of potential vorticity dynamics
and the action of counter propagating Rossby waves \citep[to name just a few]{bretherton1966propagation,hoskins1985use,baines1994mechanism,
heifetz1999counter},  and the RWI has been shown to be consistently interpreted
within that framework \citep{umurhan2010potential}.   Potential vorticity 
\footnote[2]{sometimes known in the astrophysical
literature as ``vortensity"} 
(PV) in this context
is equal to $q/\rho$ in which $q\equiv(\nabla\times{\bf u})\cdot\hat{\bf z}$ is the vorticity in
the disc vertical direction and $\rho$ is the density.   The instantaneous basic state rotation
vector when the disc is in exact rotational balance with no contribution from radial pressure
support is given by $\Omega_k(r){\bf {\hat z}}$ with $\Omega_k(r)
= \Omega_k(r_0)(r_0/r)^{3/2}$ where $r$ is radius in a cylindrical coordinate system and
where $r_0$ is a fiducial radius with corresponding rotation rate $\Omega_k(r_0)$.
As a general rule for axisymmetric basic states, for every radial extremum in the PV gradient there can exist
an azimuthally propagating 
Rossby wave radially localized around the corresponding radial extremum point. 
In the simplest configuration in which there is some action-at-a-distance and
no possibility of critical layers (cf. \citet{marcus2013three}), instability
can arise from the resonant interaction of two or more Rossby waves \citep{heifetz1999counter}.
For example, in RWI unstable discs with a single axisymmetric pressure bump, there
are two oppositely signed extrema in the radial gradient-PV which can each support Rossby waves
with wave speeds boosted by the local flow of the disc \citep{umurhan2010potential}.  
Resonant interaction and subsequent instability can occur 
when the Rossby waves move with equal and opposite directions in some reference frame. We also note that the previous studies by  \citet{mamatsashvili2009vortices} and \citet{mamatsashvili2007transient} have explored the PV profiles in order to understand  the evolution of the dynamics in PP-discs,  
\par
This perspective of interacting Rossby waves and resonant instability has been extended to include
other physical effects, e.g. the interaction of resonant
Rossby-gravity waves \citep{harnik2008buoyancy,rabinovich2011vorticity} and Alfven-Rossby waves in sheared MHD flows
\citep{heifetz2015interacting}.  Generally when there is some additional physical
effect included in shear flows, baroclinicity begins to take on an important dynamical role.  
In the case of the above-mentioned Rossby-gravity waves, pressure and density isolines of the perturbed flow do not
coincide  (i.e. $\nabla \rho \times \nabla p \neq 0$) and, consequently, 
a certain amount of vorticity creation can occur.  Places in 
a zonal disc flow in which there is a stark density gradient can support vorticity waves \footnote[3]{Although vorticity waves and Rossby waves are often used interchangeably,  throughout this manuscript we distinguish between them. We consider Rossby waves as associated with vorticity production by the advection of the mean shear, and vorticity waves as  associated with any vorticity creation.} with this baroclinic ``vorticity creation" effect and this, in turn, can influence
the other vorticity waves of the physical system and consequently change the stability character of the whole 
physical arrangement.A physical perspective developed for a self-gravitating sheared system should therefore take this
baroclinic effect into account.
\par
\citet{LH2013} have extended the examination
of the RWI to infinitely thin self-gravitating discs and they show that, among other things,
transition to instability is very subtle and that, if the conditions are right, a localized
pressure deficit (as opposed to a pressure bump) can also lead to instability. 
It is the aim of this work to 
begin constructing a potential vorticity view to use in understanding unstable 
transition in sheared self-gravitating
configurations.  Additionally,
careful numerical experiments of self-gravitating disc systems performed
by  \citet{lin2011edge, lin2011effect} \footnote[4]{We note that \citet{lin2011edge, lin2011effect} consider 2D, razor-thin disks. In the next paragraph we shortly discuss both the razor-thin and the infinite cylinders disk approximations.} examined the 3D disc response of a radial section of
a disc supporting a gap cleared out by a perturbing planetary body.
The numerical experiments show that the resulting azimuthally averaged radial PV of their discs exhibit
complicated gradient-PV structure, especially in the near vicinity of the two gap walls.  
Interestingly, only one of the gap
walls is dynamically stable while the other shows instability followed by restricted vortex roll up
(see Figure 6 of \citet{lin2011effect}).
The effort of this first of a series of studies is (also) motivated by making mechanistic 
and rational sense of these results.
\par
The inclusion of self-gravity in any fluid examination is subtle because
it introduces a certain amount of ellipticity to the underlying PDE's.  In practice this
means having to match the solutions developed within the interior of the fluid onto external
solutions with appropriate far-field boundary conditions. To circumvent
the difficulties inherent to this process and simplify the posed mathematical problem, 
most examinations of accretion
disc dynamics in the astrophysical literature treat the disc as either (i) vertically 
thin infinitesimally speaking (``razor thin") with a non-zero vertically integrated surface density
or (ii) as a vertically infinite cylinder.  The former is relatively 
tractable analytically as
far field solutions are straightforward to construct while the latter considerably simplifies 
the system because boundary conditions are needed to be specified in the radial direction only 
and these, in turn, may be represented as parameters of the basic state.
Of course, both have severe shortcomings if realism is the goal.  However,
as far as the action of localized edgewaves under the influence of self-gravity is concerned, 
using one or the other of the two approximations ought not contaminate the 
physical content of any result developed forthwith.  The model analyzed in this
study will be in the vertically infinite cylinder setting, which is basically
a two-dimensional system and the simplest of the two analytically tractable possibilities.
\par
In section 2 we formulate the equations of motion in this two-dimensional model setting and develop
an evolution equation for the disc vorticity.  We further develop
both basic states and perturbations. Section 3 concentrates on dynamics where
the basic state shows a radial jump in the vorticity and density fields.  The ensuing interfacial
vorticity wave dynamics are deconstructed into the dynamics as arising from
standard Rossby edge wave dynamics  and the concomitant presence
of interfacial gravity waves (sections 3.1 and 3.2).  
The latter of these is further analyzed in terms
of Boussinesq and non-Boussinesq contributions (sections 3.2.1 and 3.2.2).
Most importantly, it is in the latter of these subsections where
we identify the dynamics arising from self-gravity
as being in the Boussinesq and  non-Boussinesq category of responses.
These deconstructed analysis elements are reunited in section 3.3 and shown
how they effect the overall stability of the system.  It is in this section
we show how, indeed, self-gravity always has a stabilizing influence irrespective
of the radial stratification of the basic state.  In section 4 we
discuss the results and their applications for further analysis.

%
%
%
\section{Formulation}

\subsection{Disc equations}

\begin{figure}[t!]
\begin{center}  
\subfigure[The Rossby term.]{\label{fig:1a}
\resizebox*{7cm}{!}%
{\fbox{\includegraphics{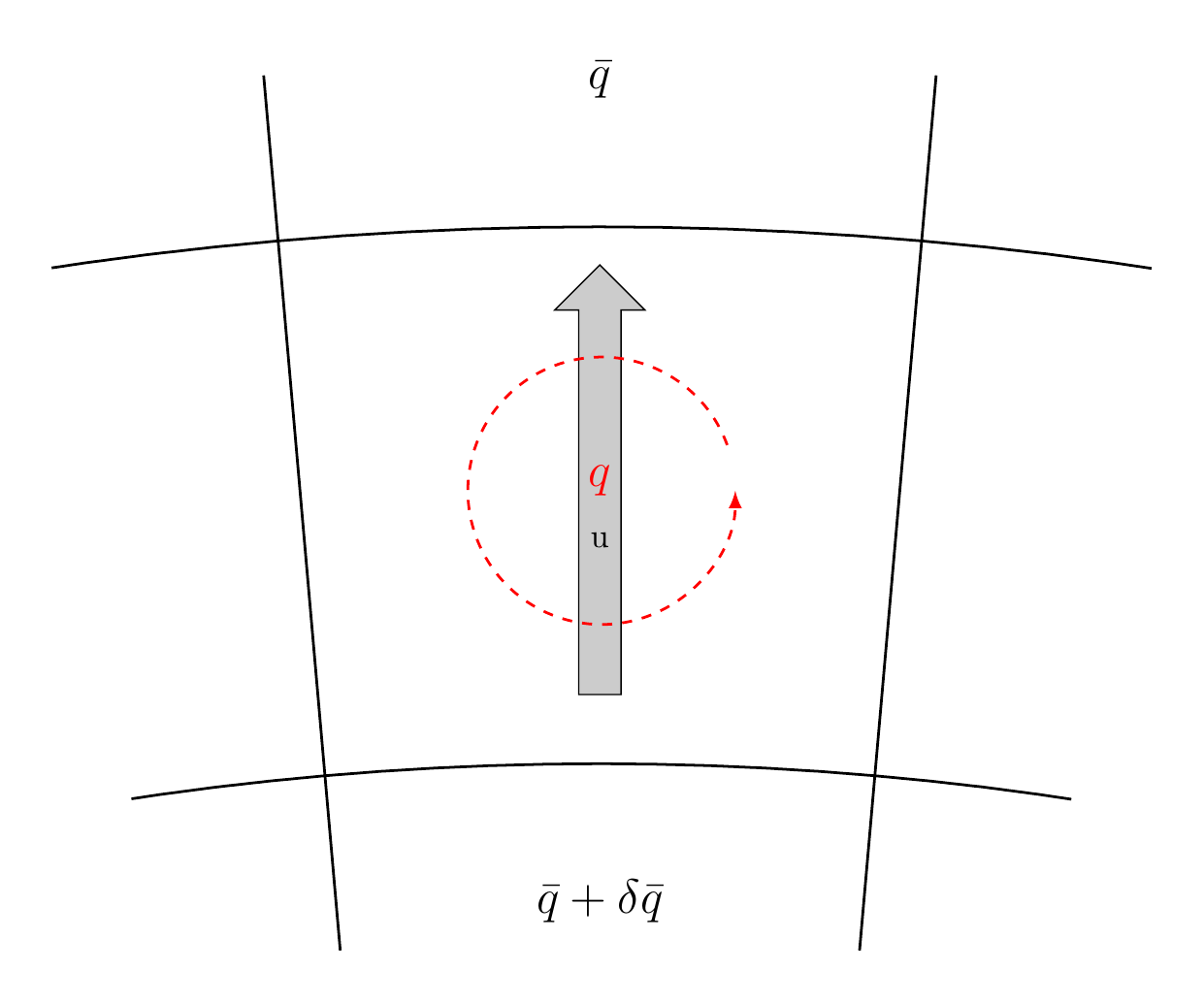}}}}
        ~ 
\quad
\subfigure[The Boussinesq term.]{\label{fig:1b}
\resizebox*{7cm}{!}%
{\fbox{\includegraphics{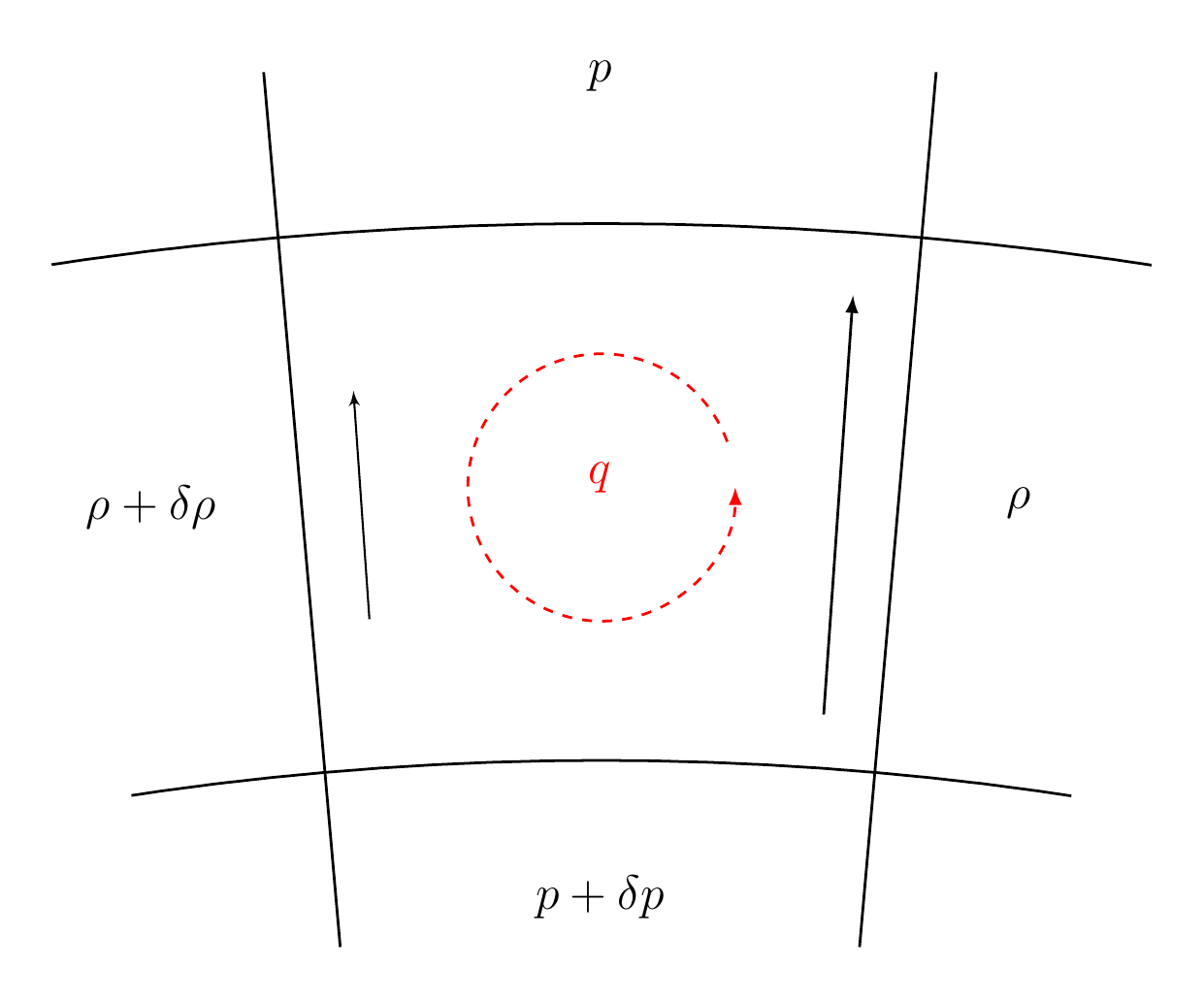}}}}

\quad
\subfigure[The non-Boussinesq term.]{\label{fig:1c}
\resizebox*{7cm}{!}%
{\fbox{\includegraphics{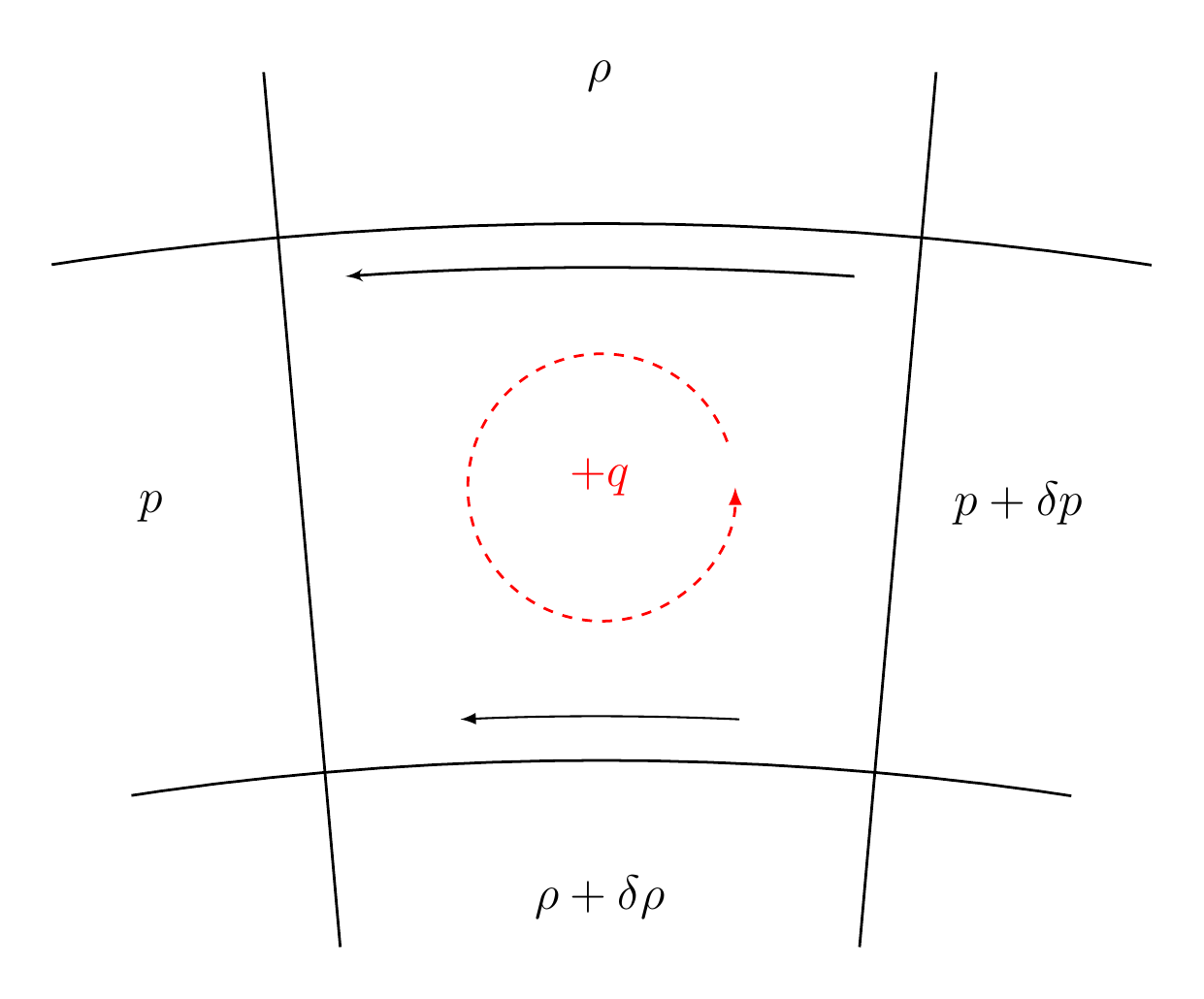}}}}
\caption{{Mechanisms of vorticity generation $q$ (indicated by a counterclockwise dashed circle) in cylindrical geometry. (a) Radial advection of the mean disc vorticity (accounted for the Rossby wave mechanism, Section 3.1). The thick arrow represents the radial velocity. (b\&c) The two components of the baroclinic torque (accounted for the Boussinesq and non-Boussinesq wave mechanisms, respectively, Section 3.2,3). Black  arrows indicate the direction and relative strength of the velocity induced by the torque. The order of the terms at the RHS of equation (7) corresponds to the order of the diagrams.}}
\label{fig:1}
\end{center}
\end{figure}

We consider a 2D gaseous disc model whose radial ($r$) and azimuthal ($\theta$) momentum equation components are written as,
\be
\label{m1_inert}
\frac{Du}{Dt}-\frac{v^2}{r}  =-\frac{1}{\rho}\frac{\partial p}{\partial
r}+\frac{\partial \phi}{\partial r} - \frac{G_{{\rm grav}}M}{r^2},
\ee
\be
\label{m2_inert}
\frac{Dv}{Dt}+\frac{uv}{r} =-\frac{1}{\rho
r}\frac{\partial p}{\partial \theta}+\frac{\partial \phi}{r\partial
\theta}.
\ee
The dynamics is described in the inertial frame of the disc where $u$ and $v$ are the
radial and azimuthal velocities of the flow.  We have included in
equation (\ref{m1_inert}) {\emph{a model Keplerian potential}} representing the gravitational
force of an object with mass $M_*$ located at $r=0$.
Our strategy is to remove this explicit potential and rewrite the equations as a deviation
around this basic rotationally supported state.  Thus the basic balance is between the full
centripetal acceleration term $v^2/r$
and the star's radial force.  Thus we write $v=v_k \equiv r\Omega_k$, where $v_k$ is
the basic Keplerian flow, so that
\beq
\frac{v_k^2}{r} = \frac{G_{{\rm grav}}M_*}{r^2}, \qquad
\Longrightarrow \quad 
{\Omega}_k=\sqrt{\frac{G_{\rm{grav}}M_*}{r^3}}.
\eeq
We then rewrite the azimuthal velocity as this Keplerian base state plus a deviation flow,
$v \mapsto v_k + v$ where $v$ is henceforth understood to be the deviation azimuthal
velocity from this basic Keplerian state.  As such, equations (\ref{m1_inert}-\ref{m2_inert})
are rewritten into the more convenient form as
\be
\label{m1}
\frac{Du}{Dt}-\left(\frac{v^2}{r} + 2\Omega_k v\right)  =-\frac{1}{\rho}\frac{\partial p}{\partial
r}+\frac{\partial \phi}{\partial r},
\ee
\be
\label{m2}
\frac{Dv}{Dt}+u\left(\frac{v}{r}+\frac{1}{r}\frac{\partial r^2\Omega_k}{\partial r}\right) =-\frac{1}{\rho
r}\frac{\partial p}{\partial \theta}+\frac{\partial \phi}{r\partial
\theta},
\ee
where the Lagrangian time derivative is given as
\beqa
\frac{D}{Dt}=\left(\frac{\partial}{\partial t} + \left({\bf u}+\Omega_k r \hat\theta\right) \cdot \nabla \right)=
\frac{\partial}{\partial t} + u\frac{\partial}{\partial r}+\left(\frac{v}{r} + \Omega_k\right)\frac{\partial}{\partial\theta}.
\nonumber 
\eeqa
where ${\bf u} \equiv u {\bf{\hat r}} + v {{\hat {\bm\theta}}}$.
In this way, the rotationally supported state is built directly into these equations by
having explicitly removed the potential arising from the central gravitating object.
\footnote[5]{In an exact treatment of a three-dimensional system rotating around
a central potential, it is a fundamental
fact (e.g. \citet{kippenhahn1990stellar}) that if the equation of state
is assumed to be barotropic then it follows that the basic rotational state
$\Omega_k$ will necessarily be independent of the vertical coordinate.  The 
model potential considered in equations (\ref{m1_inert}-\ref{m2_inert}), while not exactly correct as far
as a real disc is concerned, is meant to be a facsimile of the aforementioned
fact.}  
When we refer to the rotation rate at some position $r_0$ we write $\Omega_k(r_0) \equiv \sqrt{\frac{G_{grav}M_*}{r_0^3}}$.
The disc surface gas density is $\rho$, $p$ represents the gas pressure, and $\phi$ is the gravitational potential resulting from the gaseous material within the disc, satisfying the Poisson's relation:
\be
\label{poisson}
\nabla^2 \phi=4\pi G_{grav}\rho.
\ee
We assume further that the gas is incompressible\footnote{The inclusion of self-gravity effects both compressible and incompressible disturbances in a flow. ?or gravitationally unstable disks it is understood that compressible/acoustic dynamics lead to instability. ?e are therefore careful to consider and apply the results of this incompressible analysis only to disk conditions where $Q>1$, since compressible modes are not known to be unstable when that is the case.?urthermore, if the time scales of the dynamics are long enough (compared to the sound crossing times across the relevant scales of the system) we may safely consider incompressible/anelastic modeling and examine the effects of SG in these types of modes in isolation. ?t is then true that the effects of SG are communicated via the deformation of the interfaces - and it is this process we seek to better understand.}, hence
\be
\label{c1}
\nabla\cdot {\bf u}=\frac{1}{r}\frac{\partial}{\partial
r}(ru)+\frac{1}{r}\frac{\partial v}{\partial \theta}=0\, ,
\ee
and that, consequently, the local density of the gas is an advected passive tracer:
\be
\label{c2}
\frac{D\rho}{Dt}=0.
\ee
Equations (\ref{m1}-\ref{c2}) form a complete set to describe the evolution of the disc variables $(u,v,\rho,p,\phi)$.
In this paper however we focus on interfacial vorticity waves, thus we transform (\ref{m1}) and (\ref{m2}) into the vorticity equation:
\be
\label{vorticity1}
\frac{D}{Dt}\left[(\nabla\times{\bf u})\cdot\hat{\bf z}\right]
+ u \frac{\partial Q_{{\rm kep}}}{\partial r}
= {1\over \rho^2} \big(\nabla\rho\times\nabla p\big)\cdot\hat{\bf z} 
\ee
or more explicitly after moving the Coriolis-like term onto the righthand side of the above expression,
\be
\label{vorticity2}
\frac{Dq}{Dt}=-u\frac{\partial Q_{{\rm kep}}}{\partial r}
-\frac{1}{r\rho ^2}\left[\frac{\partial \rho}{\partial\theta} \frac{\partial p}{\partial r}-\frac{\partial\rho}{\partial r}\frac{\partial p}{\partial\theta}\right] ,   
\ee
where $q \equiv \frac{1}{r}\frac{\partial}{\partial r}(rv)-\frac{1}{r}\frac{\partial u}{\partial\theta}$, is the relative vorticity perpendicular to the disc plane. 
In arriving to equation (\ref{vorticity2}) we have made explicit use of equation (\ref{c1}). it should be noted that as the flow is incompressible, there is essentially no difference between the vorticity and the vortensity equation.  We
observe that
generation of vorticity $q$, arises from radial advection of the mean disc vorticity $Q_{{\rm kep}}$ associated with the Keplerian shear (Fig. \ref{fig:1a}), 
\beq
Q_{{\rm kep}} \equiv
\frac{1}{r}\frac{\partial(r^2\Omega_k)}{\partial r} ,
\eeq
and the baroclinic torque ${1\over \rho^2} (\nabla\rho\times\nabla p)\cdot\hat{\bf z}$, whose two terms generate shear and hence vorticity (Figs. \ref{fig:1b}\&\ref{fig:1c}). When isobars and isopicnals intersect, the same pressure gradient exerts larger acceleration on the gas whose density is smaller (simply because the pressure gradient force is $-{1\over \rho}\nabla p$).
\subsection{Basic state and linearized dynamics}

Equation set (\ref{m1}-\ref{c2}) admits an axisymmetric steady-state solutions (denoted by bars), satisfying the balance of forces:
\be
\label{base}
\frac{1}{\bar\rho}\frac{\partial \bar p}{\partial r}= \frac{\partial\bar\phi}{\partial r} +\bar v\(2\Omega_k+\frac{\bar v}{r}\),
\ee
with zero radial velocity ($\bar u = 0$).  This equation is supplemented by the solution of the Poisson equation
in steady state
\beq
\label{base_p}
\frac{1}{r}\frac{\partial}{\partial r} r\frac{\partial\bar\phi}{\partial r} = 4\pi G_{{\rm grav}} \bar\rho.
\eeq
together with appropriate boundary conditions.  The two basic state
equations are characterized
by four functions of radius, $\bar v$, $\bar \rho$, $\bar p$ and $\bar \phi$, which means
that the model as presented is unconstrained and, as a consequence,
any basic state flow profile may be specified freely by two arbitrary functions of radius.  
As we are concerned with the dynamics localized primarily in the region
of some radial position $r_0$, the values
of particular basic state quantities evaluated there, like $\bar v$ and $\frac{\partial \bar p}{\partial r}$
(which turn out to be important in what follows),
can be and are treated henceforth as parameters of the system.
\par
\bigskip
Linearization of the vorticity equation (\ref{vorticity2}) with respect to this basic state reveals:
\be
\label{v2}
\frac{D_{_L}q'}{Dt}=-u'\frac{\partial}{\partial r}\left(\bar q+\frac{1}{r}\frac{\partial(r^2\Omega_k)}{\partial r}  \right)-\frac{1}{r\bar\rho^2}\left[\frac{\partial \rho'}{\partial\theta}\frac{\partial \bar p}{\partial r} - \frac{\partial\bar\rho}{\partial r}\frac{\partial p'}{\partial\theta}   \right],
\ee
where ${\bar q} = \frac{1}{r}\frac{\partial}{\partial r}(r{\bar v})$ is the basic state vorticity, small perturbations from the basic state are denoted by primes,  and 
 $\frac{D\sub{L}}{D{t}}\equiv\frac{\partial}{\partial{t}}+\left(\frac{\bar v}{r}+\Omega_k\right)\frac{\partial}{\partial \theta}$ is the linearized Lagrangian time derivative.
Hence, vorticity perturbation may be generated by radial advection of the mean ambient (Keplerian plus basic state) 
vorticity 
\beqa
\bar Q = \bar q + Q_{{\rm kep}} = \bar q + \frac{1}{r}\frac{\partial(r^2\Omega_k)}{\partial r} ,
\nonumber
\eeqa 
and by the linearized action of the baroclinic torque, where its second term is neglected when assuming a Boussinesq flow. The baroclinic torque can be written in terms of the radial displacement perturbation $\xi'$ ($u' = \frac{D_{L}}{Dt}\xi'$) when noting that linearization of the incompressible continuity equation (\ref{c2}) implies:   
\be
\label{rho_xi}
\frac{D_{_L}{\rho'}}{Dt}=-u' \frac{\partial\bar\rho}{\partial r}\,\, \Rightarrow \,\, \rho'=-\xi'\frac{\partial\bar\rho}{\partial r},
\ee 
that is to say that all small density perturbations result from radial advection of the basic state density. Substitute (\ref{rho_xi}) into (\ref{v2}) gives: 
\be
\label{v22}
\frac{D_{_L}q'}{Dt}=-u'\frac{\partial \bar Q}{\partial r} +\frac{1}{r\bar\rho^2}\frac{\partial\bar\rho}{\partial r}
\der{}{\theta}\left[\xi'\frac{\partial \bar p}{\partial r} +p' \right].
\ee
In barotropic incompressible flow material conservation of density implies material conservation of pressure and hence, similar to (\ref{rho_xi}),\, 
$p'_{barotropic}=-\xi'\frac{\partial\bar p}{\partial r}$. Therefore, the sum of the terms in the squared brackets represent the residual baroclinic pressure perturbation.  

\section{Interfacial vorticity wave dynamics}

As we mentioned in the Introduction, both numerical experiments and observations suggest that sharp gradients of vorticity and density emerge in different stages of the disc evolution.  Such sharp gradients (either
in density or vorticity) will support the generation of interfacial (edge) waves.   Here we wish to focus on their basic mechanism of propagation and growth and, hence for simplicity, we 
assume strong gradients existing in both the density and vorticity only and
replace the corresponding gradient terms appearing by $\delta$-functions, i.e.,
\be
\frac{\partial \bar Q}{\partial r}=\Delta\bar Q\delta(r-r_0), \qquad \frac{\partial \bar\rho}{\partial r}=\Delta\bar\rho\delta(r-r_0).
\label{basic_PV_state}
\ee
In practice we are assuming that there is no vorticity production anywhere except for  $r=r_0$, This is a valid assumption only if  $ \frac{\partial \bar q}{\partial r}>>\frac{\partial Q_{{\rm kep}}}{\partial r}=-\frac{3}{4}\sqrt{\frac{G_{grav} M_*}{r^5}}$,  therefore we expected that in these cases the dynamics can be reasonably represented  by the interfacial waves  approximation (cf. analogy to MHD shear flow \citet{heifetz2015interacting}).
\par
\begin{figure}[!]
\begin{center}  
       \subfigure[]{\label{fig:2a}
\resizebox*{0.5\textwidth}{!}%
{\includegraphics{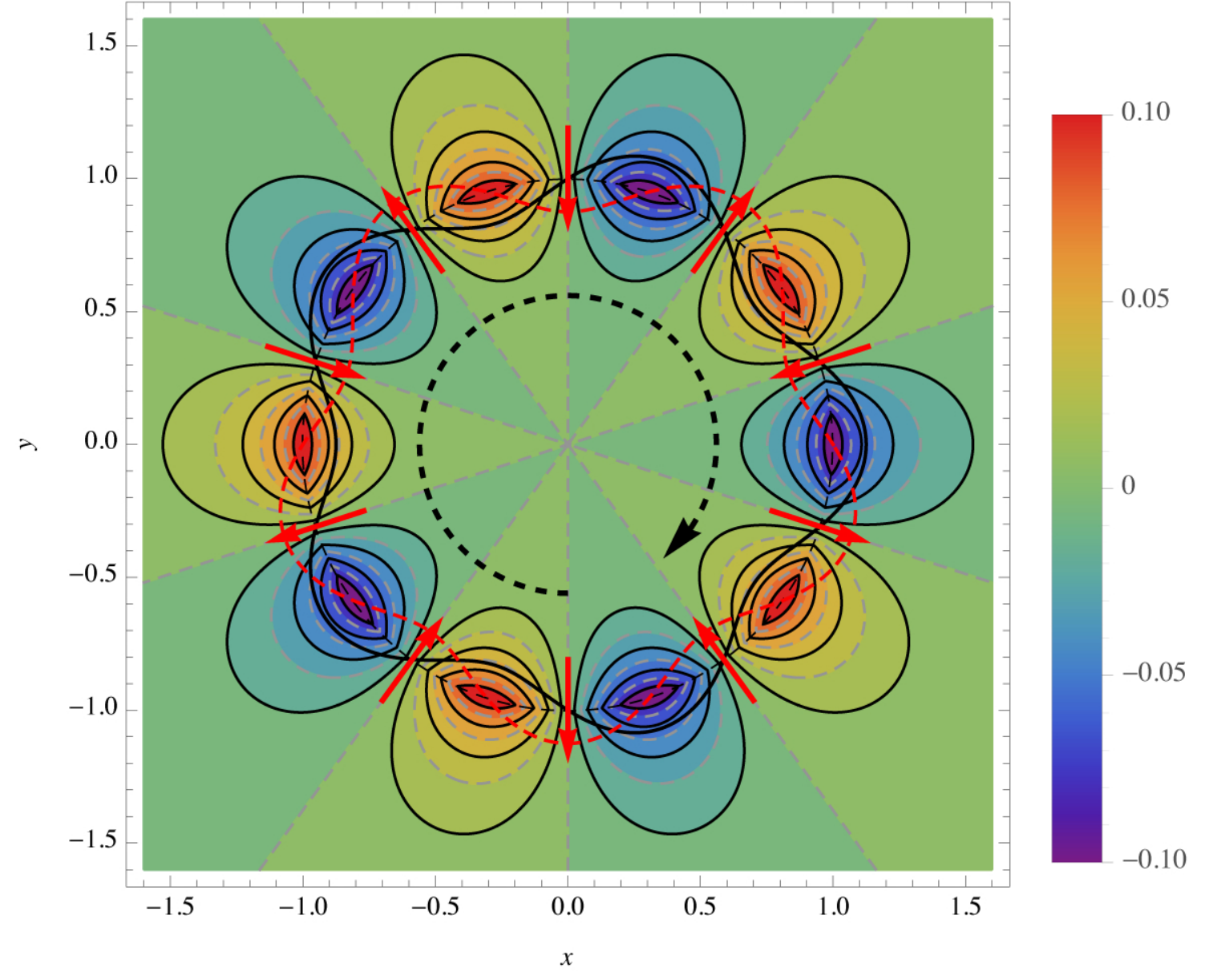}}}
        ~ 
       \subfigure[]{\label{fig:2b}
\resizebox*{0.4635\textwidth}{!}%
{\includegraphics{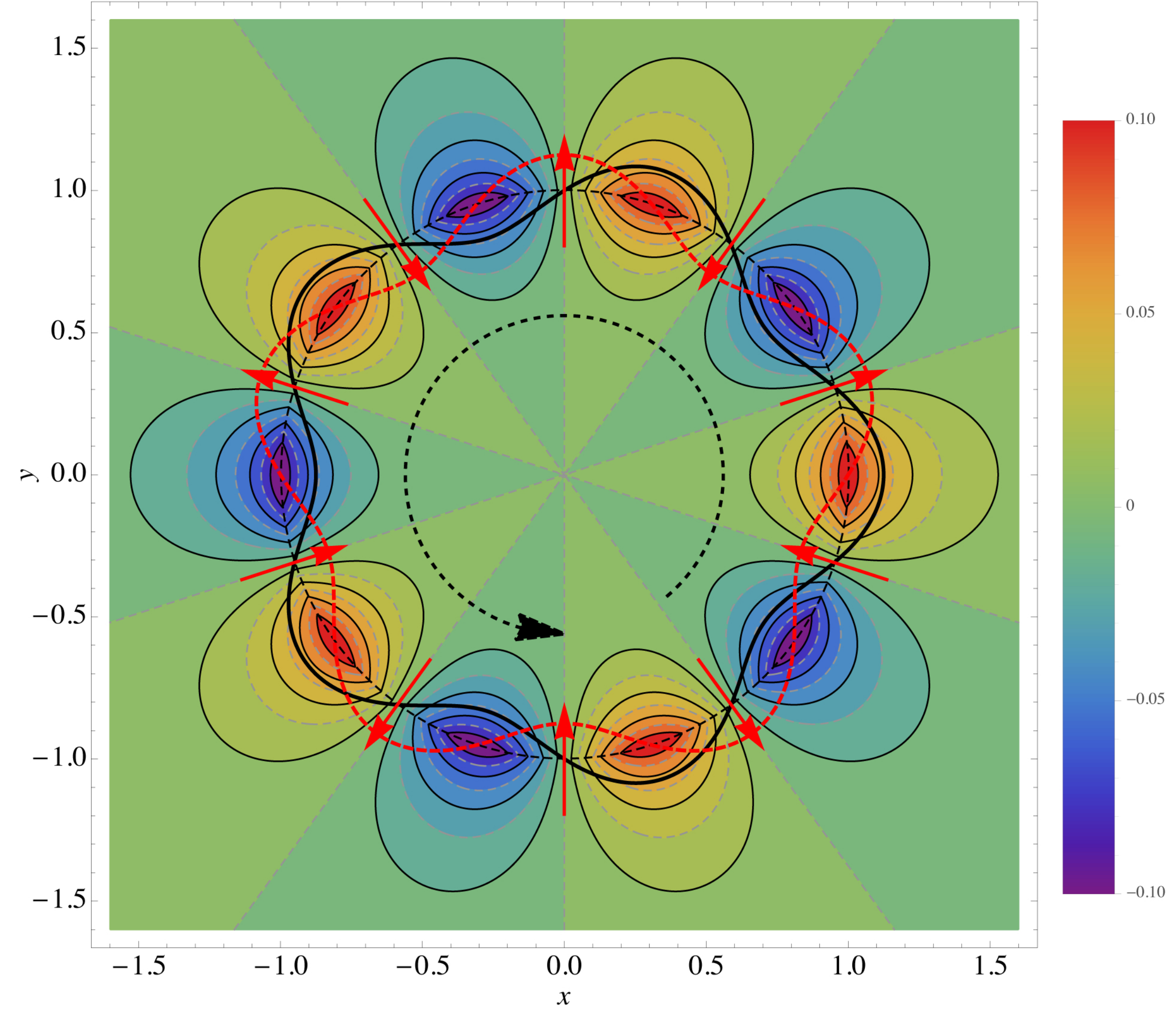}}}
        ~ 
          
\quad
\subfigure[]{\label{fig:2c}
\resizebox*{0.47\textwidth}{!}%
{\includegraphics{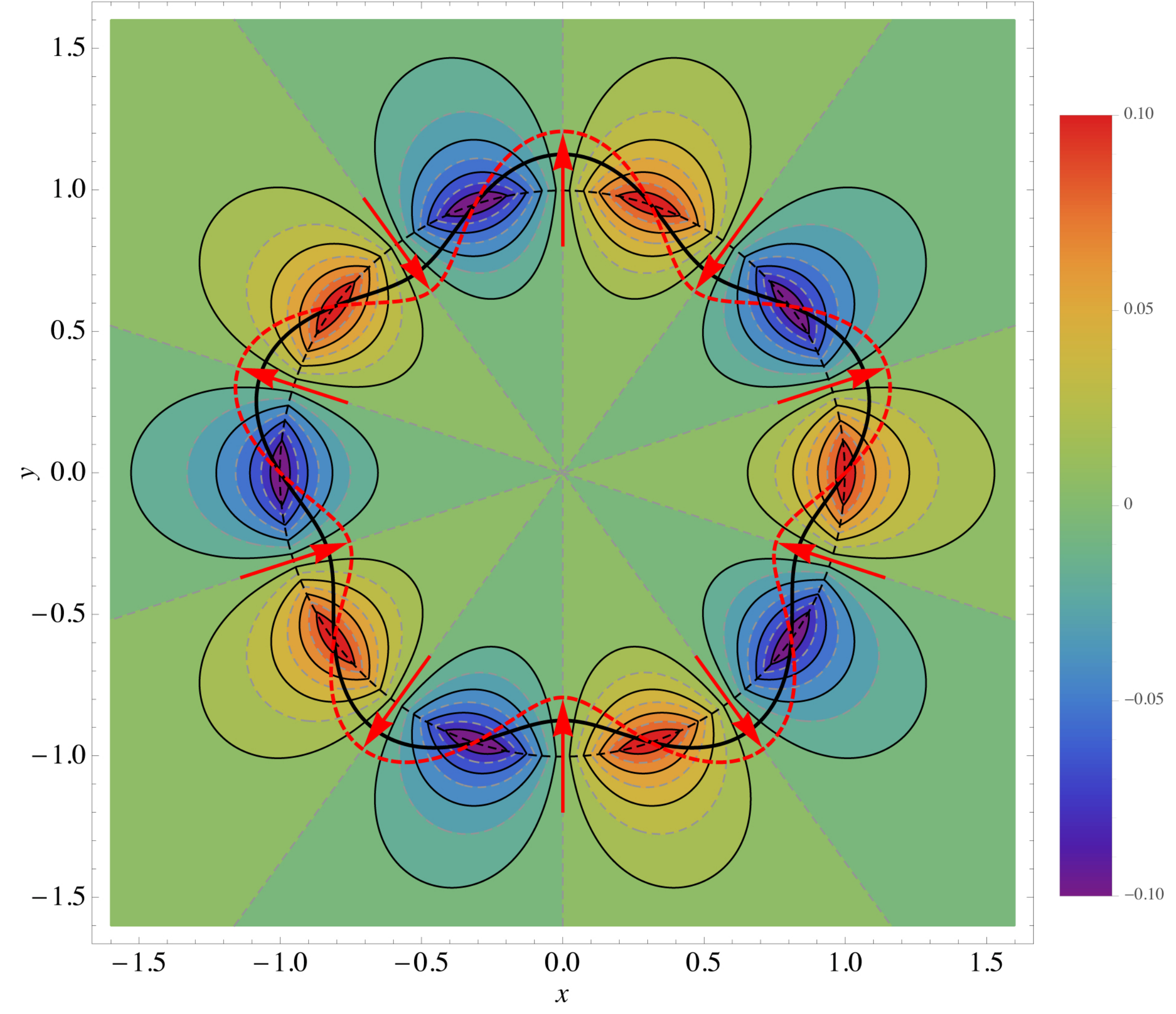}}}
        ~ 
\quad
 \caption{{Contour plots of the streamfunction $\psi'$ (16) for an azimuthal wavenumber $m=5$, centered at the interface $r_0 = 1$. Positive values of $\psi'$ correspond to negative values of $q'$, hence the red arrows indicate the direction of the radial perturbation velocity $\frac{D_{_L}\xi}{D{t}}=u'$ corresponding to 
$\psi'$.  The black wavy line represents the interface displacement $\xi'$ at present time, while the red dashed one represents its evolution at some time $t+\Delta t$ due to the advection of $u'$. Scenarios in which the vorticity and the displacement are (a) in phase resulting clockwise propagation (with respect to the mean flow) (b) in anti-phase resulting counterclockwise propagation (c) in quadrature resulting instability. In all figures the Keplerian rotation is counterclockwise.}}
\label{fig:psi}
\end{center}
\end{figure}

Equation (\ref{v2}) implies then that the generated vorticity perturbations must have $\delta$-function structures as well. Thus for a wave-like solution the vorticity perturbation takes the form:
\be
\label{q_p}
q'(r,\theta,t)=\hat q_0\delta(r-r_0)e^{i(m\theta-\omega t)}
\ee
(where the zero subscript indicates evaluation at $r_0$ and the azimuthal wavenumber $m$ is taken to be positive). Since the gas is assumed incompressible, we can use (\ref{c1}) to introduce the perturbation streamfunction $\psi'$ satisfying:
\be
\label{stream}
u'=-\frac{1}{r}\frac{\partial \psi'}{\partial\theta}; \quad v'=\frac{\partial\psi'}{\partial r}
\ee
so that $q' = \nabla^2\psi'$. Thus in order for $\psi'$ to satisfy (\ref{q_p}), it should be in the form:
\be
\label{p_p}
\psi'=\hat q_0 G(r,r_0,m)e^{i(m\theta-\omega t)}.
\ee
where the Green function $G$ (not to be confused with the Gravitational constant denoted by $G_{grav}$) is:
\be
\label{green}
    G(r,r_0,m)=-\frac{r_0}{2m} 
\begin{cases}
    \left(\frac{r}{r_0}\right)^{-m},& \text{if } r\geq r_0\\
     \left(\frac{r}{r_0}\right)^{m},& \text{if } r\leq r_0
\end{cases}
\ee
satisfying $\nabla^2 G = {1\over r}\der{}{r}(r\der{G}{r}) - ({m\over r})^2 G = \delta(r-r_0)$, together with the boundary conditions $G(r=0,\infty)=0$. 
Note that while $u'$ is continuous, $v'$ is discontinuous at $r_0$ and this infinite shear corresponds to the $\delta$-function of the vorticity perturbation there. An example for the structure of $\psi'$ is shown in Fig. \ref{fig:psi}. 

We assume a wavelike solution for all the perturbed values in the form of $f'=\hat f(r,r_0,m)e^{i(m\theta-\omega t)}$, but before solving the general dispersion relation for equation (\ref{v22}) we will split the problem to the dynamics arises from the existence of the mean vorticity and the mean density gradients.

\subsection{Interfacial Rossby waves}

\begin{figure}[p]
\begin{center}
 \subfigure[]{\label{fig:3a}
\resizebox*{0.85\textwidth}{!}%
{\fbox{\includegraphics{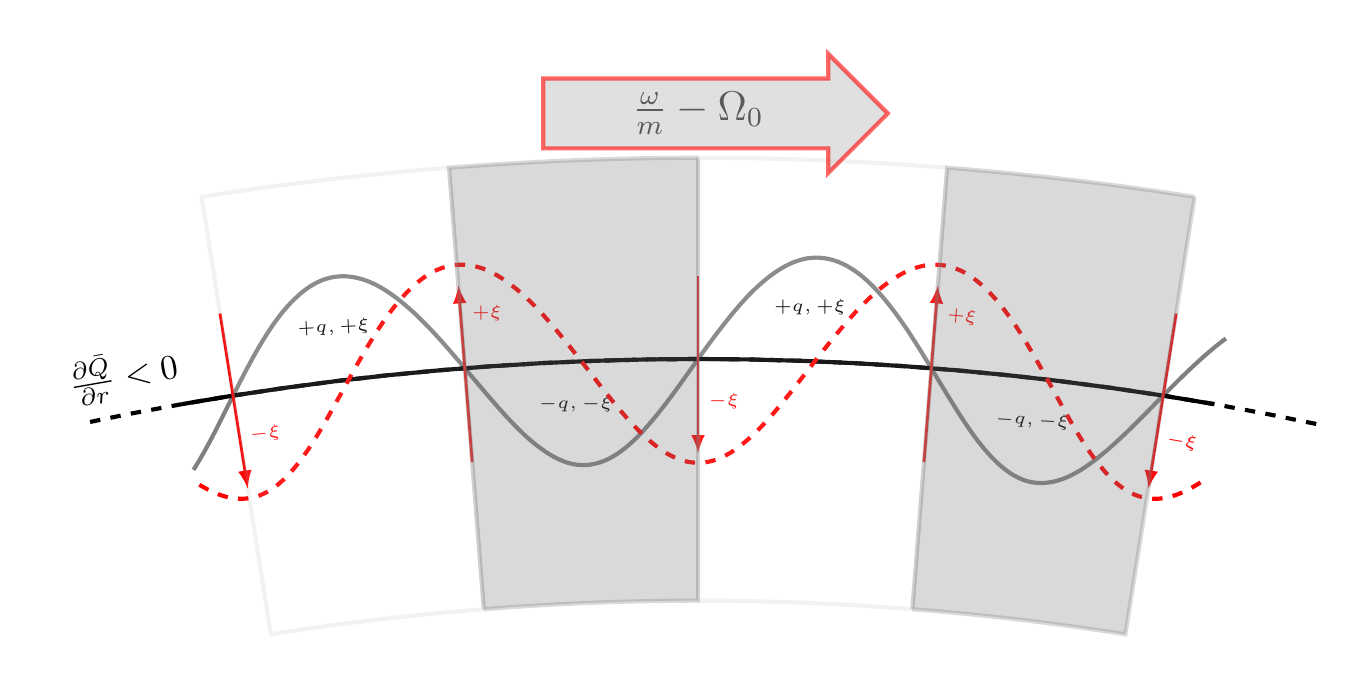}}}}    

\subfigure[]{\label{fig:3b}
\resizebox*{0.85\textwidth}{!}%
{\fbox{\includegraphics{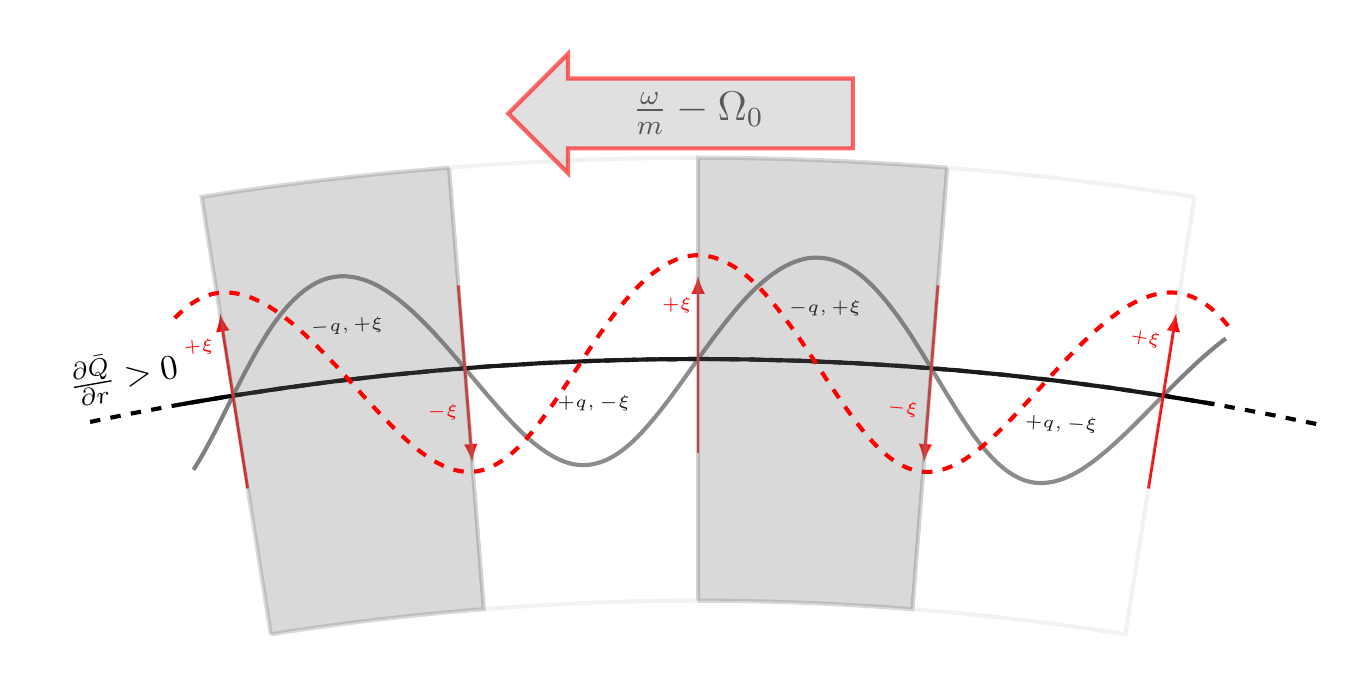}}}}    

\subfigure[]{\label{fig:3c}
\resizebox*{0.85\textwidth}{!}%
{\fbox{\includegraphics{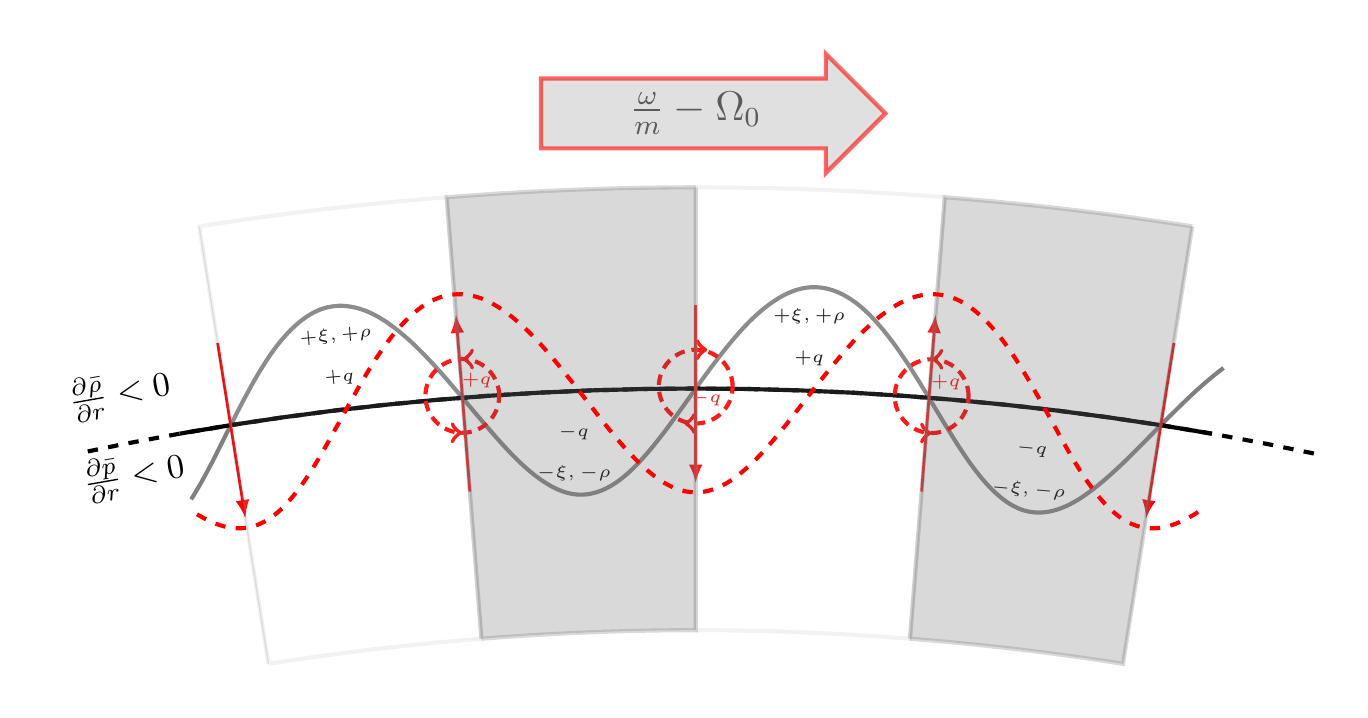}}}} 
   \end{center}
\end{figure}

 Only the basic state shear may account to sharp vorticity gradients.
In the absence of mean density gradient equation (\ref{v22}), with the basic state of (\ref{basic_PV_state}), becomes:
\be
\frac{D_{_L}q'}{Dt} = -u'\Delta\bar Q\delta(r-r_0),
\label{simple_rossby_dynamics}
\ee
 which together with (\ref{q_p}-\ref{green}), provides the dispersion relation:
\be
\({\omega_{Ros}\over m} - \Omega\sub 0\) = \({\Delta\bar Q\over 2m}\)_{r_0},
\qquad \Omega\sub 0 \equiv {\bar v_0\over r_0}+\Omega\sub k(r\sub 0),
\label{simple_rossby_wave}
\ee
where $\Omega\sub 0$ is the local rotation rate of the disc at position $r_0$.
Equation (\ref{simple_rossby_dynamics}) also implies that $\hat q_0 = -\Delta\bar Q \xi_0$, that is to say, that all variations in the vorticity result from advection of the mean vorticity   
(Fig. \ref{fig:1a}) at the interface of $r_0$. The result is an interfacial Rossby wave which propagates (relative to the mean flow) in a direction that keeps the lower mean vorticity to its left. Hence if, for instance, the mean flow and the mean vorticity gradient are of opposite signs the waves propagate counter the mean flow. The mechanism of propagation is such that vorticity anomalies induce circulation that both undulates the interface and advects the mean vorticity to generate fresh vorticity anomalies in concert with the undulated interface  (Fig. \ref{fig:3a}\&\ref{fig:3b}). These Rossby waves are the interfacial versions of the plane-like Rossby waves
\citep{rossby1939relation}  that are generic in the mid-latitudinal atmosphere due to the meridional gradient of the Coriolis force there. 
\par
The dispersion relation of (\ref{simple_rossby_wave})  indicates that the waves are neutral on their own. However two remote interfacial Rossby waves can generate instability (e.g. \citet{baines1994mechanism,heifetz1999counter}) by ``action-at-a-distance'' resonance if the necessary conditions for shear instability are satisfied, i.e., that each wave propagates counter the mean flow \citep{fjortoft1953changes} and that the radial mean vorticity gradients at the two interfaces are of opposite sign 
\citep{rayleigh1879stability}. The essence of this resonance mechanism is illustrated in (Fig. \ref{fig:4b}) and
has been applied in order to explain the RWI in \citet{umurhan2010potential}.  When self-gravity is included,
instability will manifest itself along similar lines with the same main requirement - that there
are at least two separate interfaces able to resonate with one another.  Self-gravity
as a dynamical influence only becomes relevant when there is at least one
density gradient placed somewhere in the flow.
The dynamical mechanisms resulting from such a single density gradient remains to be elucidated and this
is what we do next.

\subsection{Interfacial gravity waves}

In the absence of mean vorticity gradient together with the basic state 
expressed in equation (\ref{basic_PV_state}), equation (\ref{v22}) becomes:
\be
\frac{D_{_L}q'}{Dt} = 
\der{}{\theta}\left[\xi'\frac{\partial \bar p}{\partial r} +p' \right]
\frac{\Delta\bar\rho}{r\bar\rho^2}\delta(r-r_0),
\label{PV_equation_with_baroclinic_terms}
\ee
where $\bar\rho_0$ is evaluated as the averaged mean density across the jump  . We first obtain the solution for an interfacial gravity wave under the Boussinesq approximation which neglects the pressure perturbation in the squared brackets. 
   
\subsubsection{Boussinesq waves}
At the position $r=r_0$ the cross-stream displacement equation $u' = \frac{D_{L}}{Dt}\xi'$, together with the vorticity equation there yields:
\be
\hat q_0  = -2m\({\omega \over m}-\Omega\sub 0\){\hat \xi}_0, \qquad 
\({\omega \over m}-\Omega\sub 0\)\hat q_0 = -\(\frac{\Delta\bar\rho}{r\bar\rho^2}\frac{\partial \bar p}{\partial r}\)_{r_{0}}\hat \xi_0,
\label{2_equations_with_Boussinsesq}
\ee
and the corresponding dispersion relation is:
\be
{\omega\sub{{{\rm {Bous}}}}\over m} - \Omega\sub 0 = \pm \[{1 \over 2rm}\frac{\Delta\bar\rho}{\bar \rho}\({1\over \bar \rho}\frac{\partial \bar p}{\partial r}\)\]^{1/2}_{r_0}.
\ee
The wave is neutral when the signs of ${\Delta\bar\rho}$ and  $\frac{\partial \bar p}{\partial r}$ agree. For instance, if the mean flow is relatively weak, (\ref{base}) is close to hydrostatic balance and under stable stratification (${\Delta\bar\rho}<0$) $\frac{\partial \bar p}{\partial r}<0$. The wave propagation mechanism for such scenario is illustrated in Figs. 3(c,d). For clockwise propagation relative to the mean flow, 
$\displaystyle \({\omega\sub{{{\rm {Bous}}}}\over m}-\Omega\sub 0\)<0$, equation set (\ref{2_equations_with_Boussinsesq}) determines that $q'$ and $\xi'$ are in phase, hence the radial velocity field (indicated by the radial arrows) induced by the vorticity perturbation shifts the displacement in the clockwise direction. Since ${\Delta\bar\rho}<0$, (\ref{rho_xi}) indicates that positive radial displacements corresponds to  positive density anomalies, thus for this wave $q'$ and $\rho'$ are in phase as well. Now the Boussinesq part of the baroclinic torque results from the first term at the squared brackets of (\ref{v2}) and according to the mechanism described in Fig. \ref{fig:2a} this will shift the vorticity anomalies clockwise, in concert with the radial displacement. For the counter-clockwise solution the vorticity and the displacement anomalies are in anti-phase where the displacement and the density anomalies remain in phase. 

The unstable stratified case (${\Delta\bar\rho}>0$ \& $\frac{\partial \bar p}{\partial r}<0$) generates instability of the Rayleigh-Taylor type. The wave is  advected by the mean flow $\bar v_0$ 
and grows with a rate given by
 \beq
 \Bigg \{
 \Big|\frac{m}{2r}\Big|\frac{\Delta\bar\rho}{\bar \rho}
 \left(\frac{1}{\bar \rho}\frac{\partial \bar p}{\partial r}\right)\Bigg\}^{1/2}_{r_0}.
 \eeq
 The perturbation potential vorticity $q'$ is shifted by a quarter of a wavelength counter-clockwise to $\xi'$ and $\rho'$ and $\xi'$ are in anti-phase (Fig. \ref{fig:4a}). The resonance amplification results from the fact that in this configuration $u'$ is in phase with $\xi'$, and the Boussinesq baroclinic torque is in phase with 
$q'$.  

We thus see that both the propagation rate and the instability mechanism of Boussinesq gravity waves depend heavily on the sign of the mean radial pressure gradient force. Hence, it is crucial to include the contribution of the mean self gravity in (\ref{base}) to the calculations, if the mean potential gravity gradient is comparable in magnitude to the Coriolis and the centrifugal forces.    

\begin{figure}[p]
\begin{center}
\subfigure[]{\label{fig:3d}
\resizebox*{0.85\textwidth}{!}%
{\fbox{\includegraphics{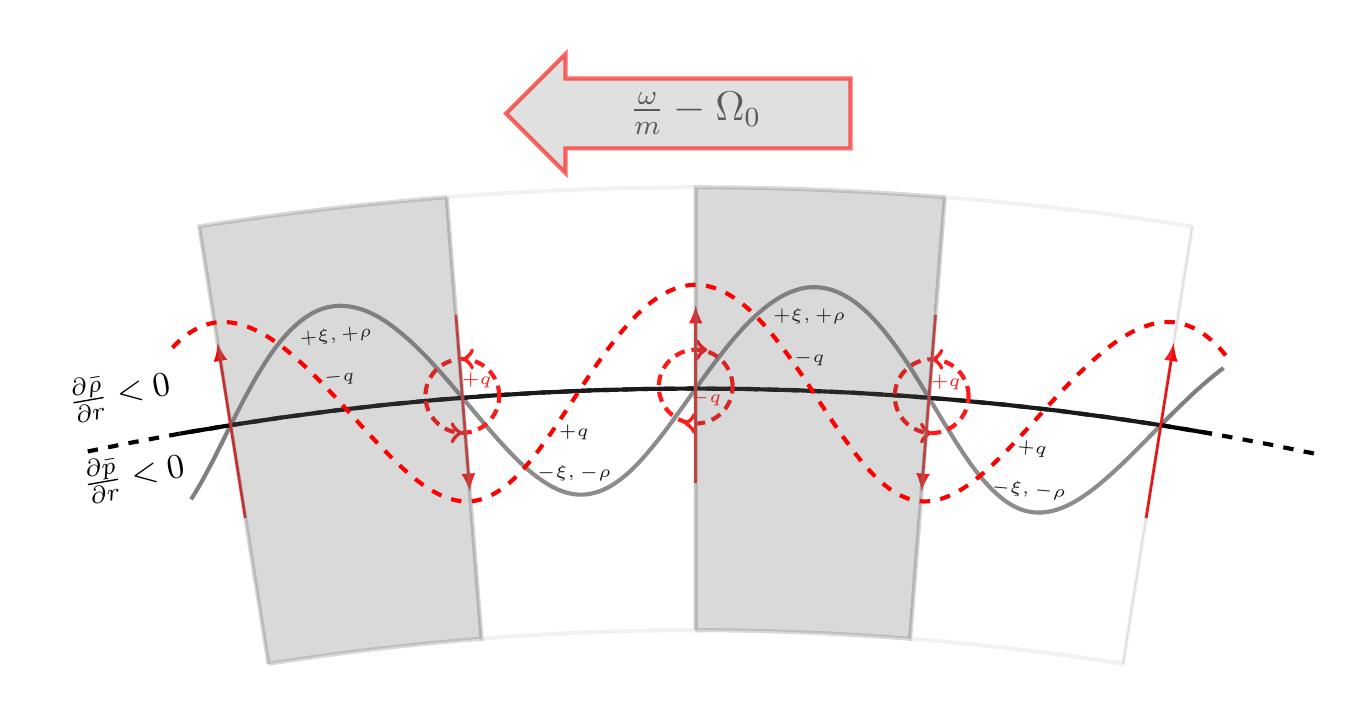}}}}    
   
\subfigure[]{\label{fig:3e}
\resizebox*{0.85\textwidth}{!}%
{\fbox{\includegraphics{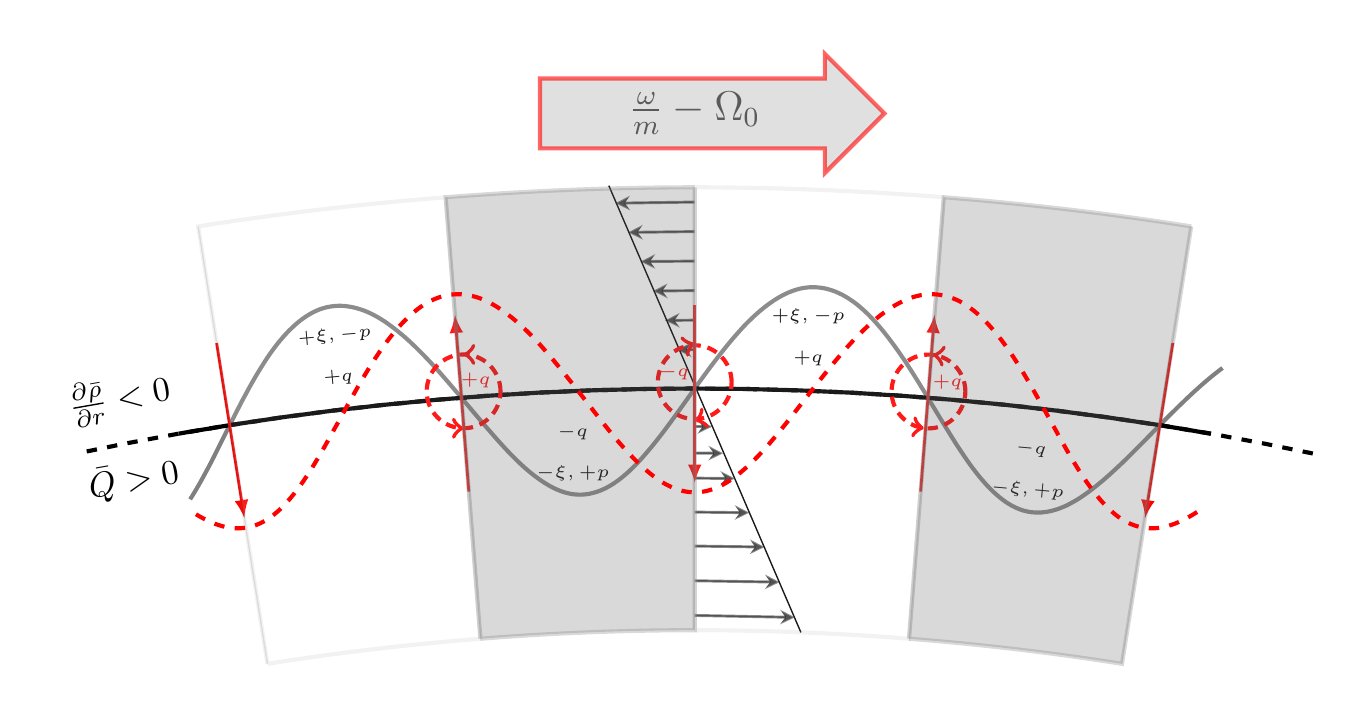}}}}  
  
\subfigure[]{\label{fig:3f}
\resizebox*{0.85\textwidth}{!}%
{\fbox{\includegraphics{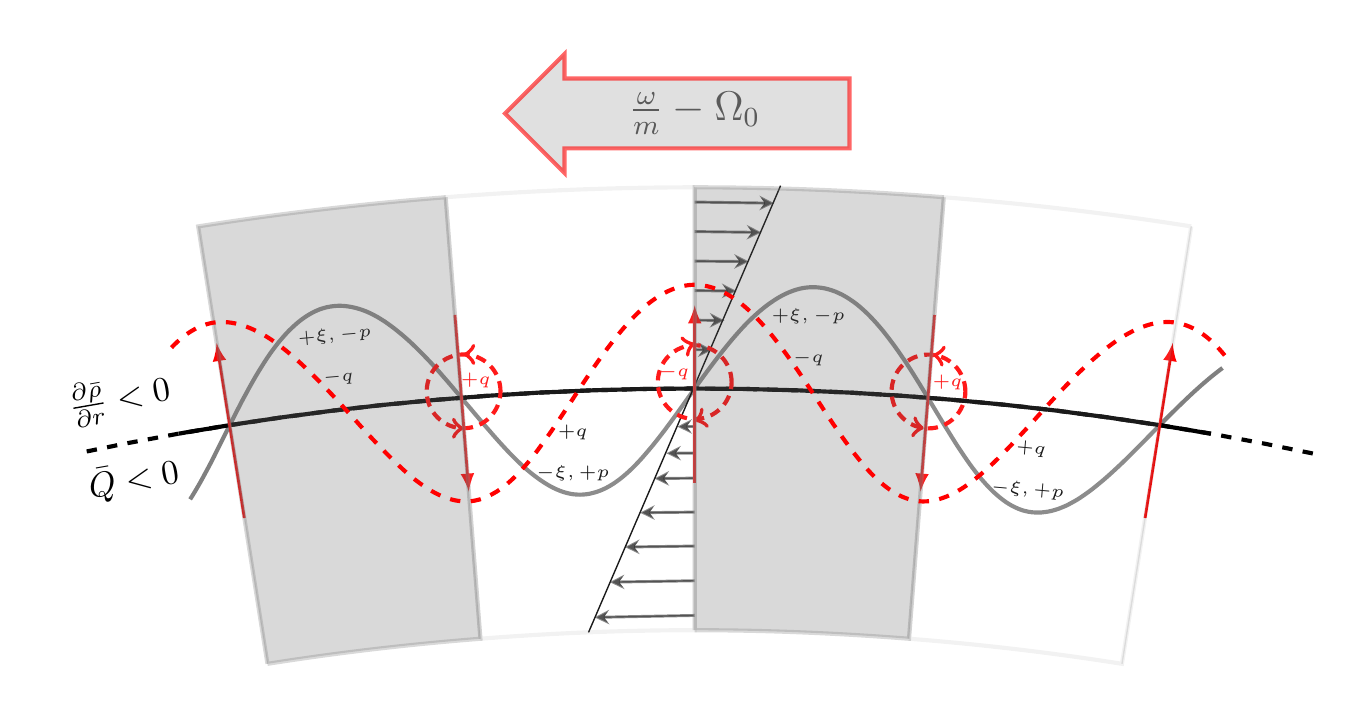}}}}  
 \end{center}
\end{figure}

\begin{figure}[p]
\begin{center}
\subfigure[]{\label{fig:3g}
\resizebox*{0.85\textwidth}{!}%
{\fbox{\includegraphics{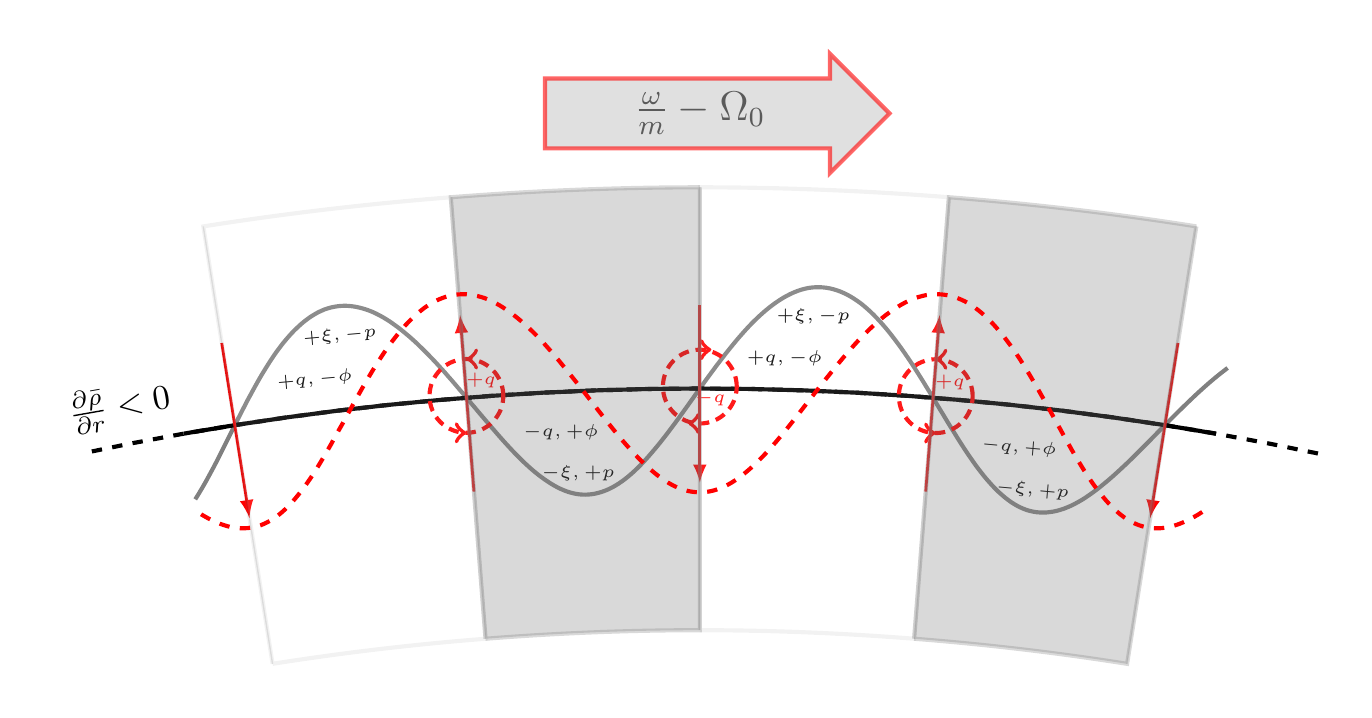}}}}  
 
\subfigure[]{\label{fig:3h}
\resizebox*{0.85\textwidth}{!}%
{\fbox{\includegraphics{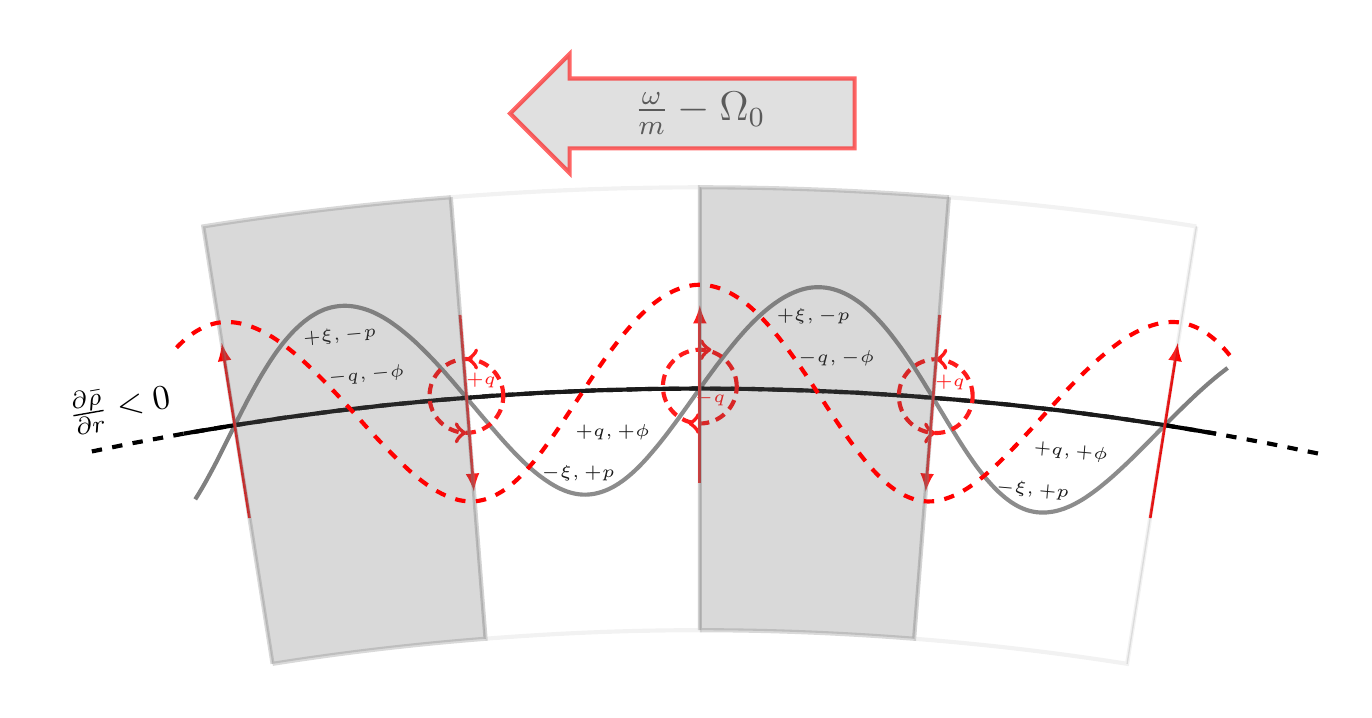}}}}  
           \caption{{Vorticity waves propagation mechanisms. The black wavy lines represent present interfacial displacements while the red dashed ones represent their evolution (as in Fig. 2).  On odd (even) figures  the waves are propagating clockwise (counterclockwise), where each pair of clockwise and anti clockwise figure corresponds to a different wave propagation mechanism: (a,b) Rossby, (c,d) Boussinesq gravity, (e,f) Non-Boussinesq gravity, (g,h) Non-Boussinesq self gravitating waves. In (a)-(h) the background white (gray) color indicate positive (negative) signs of $q'$. The red dashed circular double head arrows indicate the vorticity production. Note that the Keplerian rotation is counterclockwise, more details about the different propagation mechanisms appear in Section 3.}}%
            \label{fig:wp}
           \end{center}
\end{figure}
  
\subsubsection{Non-Boussinesq waves}

Considering only the non-Boussinesq component of the baroclinic torque, equation (\ref{PV_equation_with_baroclinic_terms}) yields:
\be
\label{nb0}
\frac{D_{_L}q'}{Dt} = 
\der{p'}{\theta}
\frac{\Delta\bar\rho}{r\bar\rho^2}\delta(r-r_0,)\qquad \Longrightarrow  \qquad 
\({\omega \over m}-\Omega\sub 0\)\hat q_0 = -\(\frac{\Delta\bar\rho}{r\bar\rho^2}\)_{r_{0}}\hat p_0\, .
\ee
In order to evaluate $\hat p_0$ we first linearize the azimuthal component of the momentum equation (\ref{m2}) to obtain:
\be
\label{m22}
\frac{D_{_L}v'}{Dt} + \bar Qu' = -{1\over r}\der{}{\theta}\({p'\over \bar \rho} -\phi'\),
\ee
and note that since $v'$ is discontinuous and changes sign at $r_0$, $\hat v_0$ must remains zero at all times.  
Hence at $r_0$ the wave satisfies the angular momentum balance between the radial flux of the absolute vorticity and the perturbation azimuthal torque:
\be
r_0\bar Q_0 u'_{0} = -\der{}{\theta}\({p'\over \bar \rho} - \phi'\)_{r_{0}},
\label{perturbation_azimuthal_torque_balance}
\ee
which with the aid of (\ref{stream}) becomes:
\be
\hat p_0 = \bar \rho_0(\bar Q_0\hat \psi_0 + \hat \phi_0).
\label{pressure_azimuthal_torque_balance}
\ee
In the absence of self gravity perturbations ($\hat \phi_0 = 0$), we can substitute $\hat \psi_0$ directly in (\ref{PV_equation_with_baroclinic_terms}) to obtain, together with the displacement stream function relation, the dispersion relation:
\be
\label{nbdr1}
{\omega\over m} = \Omega \sub 0 + \(\frac{\Delta\bar\rho}{2m\bar \rho}\bar Q\)_{r_0}\, .
\ee 
Consider for instance the stably stratified case where ${\Delta\bar\rho}<0$ and $\bar Q > 0$ (the sign of the absolute vorticity matches the sign on the Keplerian one) so that the wave is neutral and propagates in the clockwise direction relative to the mean flow (Fig. \ref{fig:3e}). Since the stream-function and the vorticity perturbations are in anti phase at $r_0$, equation (\ref{pressure_azimuthal_torque_balance})
indicates that the pressure and the vorticity perturbations are in anti-phase there. Then the non-Boussinesq component of the baroclinic torque (proportional to $-\der{\bar \rho}{r}\der{p'}{\theta}$) will generates fresh vorticity anomalies quarter of wavelength in the clockwise direction. Since the displacement $\xi'$ should be translated in concert with the vorticity this implies that it should be in phase with the vorticity anomaly. We note that this wave cannot become unstable, even in unstably stratified setup. For counterclockwise propagation 
(Fig. \ref{fig:3f}) ${\Delta\bar\rho}<0$ and  $\bar Q < 0$ $\xi'$ is in anti phase with $q'$ and $p'$. 
\par
In order to isolate the contribution of the self gravity perturbation we may assume a quasi-hydorstatic balance;\, $\hat p_0 = \bar \rho_0 \hat \phi_0$, in equation
(\ref{pressure_azimuthal_torque_balance}). 
Combining then (\ref{poisson}) with (\ref{q_p}) and  the Green's function solution gives $\nabla^2 \phi' = -4\pi G_{grav}\Delta\bar \rho \delta(r-r_0)\xi'$, hence
\beqa
\label{green2}
\phi' &=& -4\pi G_{grav}\Delta\bar \rho_0 G(r,r_0,m)\hat {\xi}_0 e^{i(m\theta-\omega t)}\, 
\nonumber \\
& & \Rightarrow  \,\,\, \hat p_0 = \bar \rho_0 \hat \phi_0 =
\({2\pi  \over m} r_0 G_{grav}\bar\rho_0\Delta\bar \rho_0\) \hat {\xi}_0.
\eeqa  
The last relation manifests the quasi-hydrostatic balance; if, for instance  ${\Delta\bar\rho_0}<0$, an outward radial displacement shifts mass outward and therefore decreases both the gravitational potential and the pressure perturbation that balances the gravity fluctuation. Substitute  (\ref{green2}) in (\ref{PV_equation_with_baroclinic_terms}), together with the generic relation obtained from the cross-stream displacement equation $u' = \frac{D_{L}}{Dt}\xi'$, we obtain:
\be
\hat q_0  = -2m\({\omega \over m}-\Omega\sub 0\){\hat \xi}_0,  \qquad 
\({\omega \over m}-\Omega\sub 0\)\hat q_0 = -{2\pi  \over m}G_{grav}\[\frac{(\Delta\bar\rho)^2}{\bar\rho}\]_{r_{0}}\hat \xi_0,
\ee
with the corresponding dispersion relation:
\be
\label{nbdr2}
{\omega\over m} = \Omega\sub 0 \pm \[{\pi \over m^2}G_{grav}\frac{(\Delta\bar\rho)^2}{\bar\rho}\]^{1/2}_{r_{0}}.
\ee
It is not surprising that quasi-hydrostatic fluctuations cannot generate instability. Consider again a relative clockwise propagation (Fig. \ref{fig:3g}), then when 
${\Delta\bar\rho}<0$ at $r_0$, $(q',\xi', \rho')$ are in phase with each other and in anti phase with $(\phi',p')$ there, ensuring the coherent propagation of the wave.
Even in the unstable stratified case (${\Delta\bar\rho}_0 > 0$), the wave remains neutral and propagates clockwise if $(q',\xi',\phi',p')$ are in phase with each other and in anti phase with 
$\rho'$ at $r_0$. For counterclockwise propagation (Fig. \ref{fig:3h}) $\xi'$ is in anti phase with $(q', \phi', p')$.

\begin{figure}[t!]
\begin{center}  
        \subfigure[SG modified RayleighÐTaylor instability]{\label{fig:4a}
\resizebox*{\textwidth}{!}%
{\fbox{\includegraphics{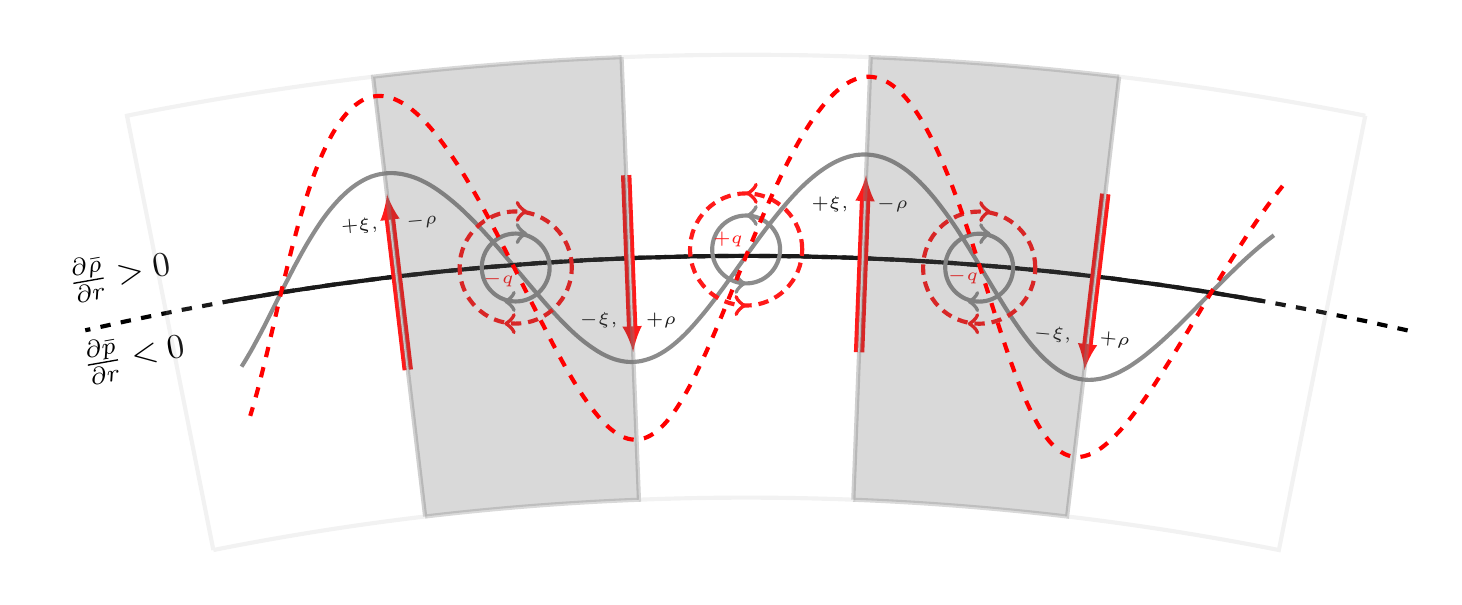}}}}    

        \subfigure[2 interfacial Rossby waves action-at-a-distance instability]{\label{fig:4b}
\resizebox*{\textwidth}{!}%
{\fbox{\includegraphics{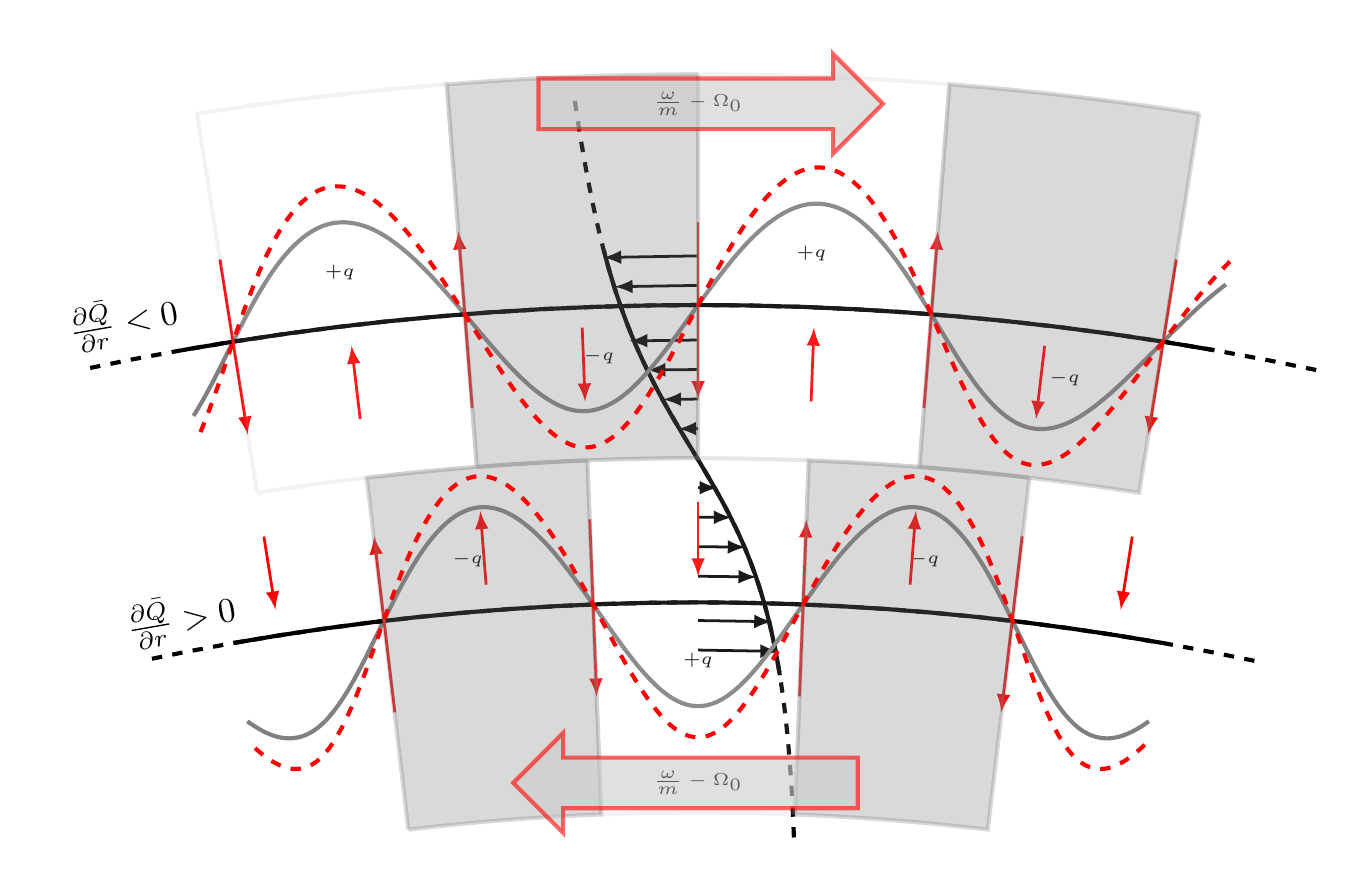}}}} 
         \caption{{Instability mechanism of interfacial vorticity waves. (a) Single interface - when the vorticity (gray circles) and the displacement (wavy gray line) are in quadrature radial velocity induced by the vorticity (doubled red arrows) amplify the displacement (dashed red wavy line) and the baroclinic torque amplify back the vorticity (red dashed circles) resulting in positive feedback. (b) An illustration of a shear flow which can support two counter-propagating vorticity wave resonance. Lines and arrows are as in the previous figures. Each wave propagates counter its local mean flow and therefore the waves may become phase locked. Furthermore, the radial velocity induced by each wave on the opposed one (indicated by the smaller evanescent red arrows) amplify the other wave's amplitude. As a result continuous mutual amplification of the waves' amplitudes is obtained. As before, the Keplerian rotation is counterclockwise.}}
\label{fig:instability}
\end{center} 
\end{figure}
\subsection{Mixed Rossby-gravity interfacial waves}

We can now combine the effects of all four components (Rossby, Boussinesq, the non-Boussineq vorticity flux and the non-Boussineq self gravity perturbation) to derive the general interfacial dispersion relation of (\ref{v22}), for the sharp gradients of equation (\ref{basic_PV_state}). We apply the same strategy applied
previously following all the substitutions we made in subsections 
(3.1,2), followed by simplifying equations (\ref{stream}-\ref{green}) to derive an expression for $\hat u_0$, and further combining
and simplifying
equations (30, 32) to obtain an expression of $\hat p_0$.  
These two terms are both expressed in terms of the mean flow properties and $(\hat q_0, {\hat \xi}_0)$. 
This process results the generalized dispersion relation:
\beqa
& & \({\omega \over m}-\Omega\sub 0\) =\frac{1}{4m}\left(\Delta\bar Q+\frac{\Delta\bar\rho}{\bar\rho}\bar Q\right)_{r_0}
\nonumber \\
& & \ \ \pm\left\{{\left[\frac{1}{4m}\left(\Delta\bar Q+\frac{\Delta\bar\rho}{\bar\rho}\bar Q\right)\right]^2+\frac{\Delta\bar\rho}{\bar\rho}\left[\frac{1}{2mr}\left(\frac{1}{\bar \rho}\frac{\partial \bar p}{\partial r}\right)_{r_0}+\frac{\pi G_{grav}\Delta\bar\rho}{m^2} \right]}\right\}^{1/2}_{r_{0}},
\label{total_dispersion_relationship}
\eeqa
together with the generic vorticity displacement ratio:
\be
\hat q_0  = -2m\({\omega \over m}-{\Omega\sub 0 }\){\hat \xi}_0\, . 
\ee
All the restoring mechanisms which, on their own, support neutral propagating waves now acting together to stabilize the only Rayleigh-Taylor like Boussinesq instability
which is obtained when the product ${\Delta\bar\rho}\cdot \frac{\partial \bar p}{\partial r} < 0$.    
Therefore, the necessary criterion for instability is the same as for 
standard Rayleigh-Taylor instability
\beqa
\({\Delta\bar\rho}\frac{\partial \bar p}{\partial r}\)_{r_0} < 0. \nonumber
\eeqa 
With the inclusion of self-gravity, the sufficient condition
for instability for a single edge wave is
\be
\left({\Delta\bar\rho}\frac{\partial \bar p}{\partial r}\right)_{r_0} < -{2r_0\over m}\[ \left({{\bar\rho}\Delta\bar Q+{\Delta\bar\rho}\bar Q \over 4}\right)^2
+ {\pi G_{grav}\bar\rho\Delta\bar\rho}^2\]_{r_0}
\, .
\label{sufficient_criterion_for_instability}
\ee 
The total composite expression within the square brackets on the right hand side of equation 
(\ref{sufficient_criterion_for_instability}) is always greater than zero, thus the
righthand side is always less than zero implying that these non-Boussinesq terms act in concert
to stabilize the familiar Rayleigh-Taylor instability. 

\section{Discussion}
This single-interface setup and analysis has been meant to be one that showcases 
the propagation and instability of individual vorticity waves 
under the influence of mean vorticity and density gradients in the presence of self-gravity. The direction of propagation and 
the ability of the wave to be unstable boils down to the phasing relationship between the vorticity and the radial displacement anomalies. When the vorticity and the displacement anomalies are in phase (antiphase), the result will be a clockwise (anti-clockwise) propagating wave as in fig \ref{fig:2a}(\ref{fig:2b}), when the displacement is in a quarter of a wavelength behind the vorticity, the wave will become unstable as in fig \ref{fig:2c}.
It is not surprising that where there was only one edge wave present
in the absence of gravitational dynamics, upon its inclusion
(both through the global pressure gradient and self-gravity)
an edge can support two so-called {\emph{kernal-gravity waves}} \citep{harnik2008buoyancy}.
It is also not surprising that
one recovers the necessary condition for Rayleigh-Taylor instability
if the flow is unstably stratified 
in the classical sense, i.e. if $\displaystyle \Delta\bar\rho \frac{\partial \bar p}{\partial r} < 0$.
\par
However, it is surprising that the 
stability criterion found in equation (\ref{sufficient_criterion_for_instability}) implies
 that in the single interface model kernel gravity waves are
 directly stabilized by self-gravity and its stabilizing role is unaffected by the
 stratification of the background state.  
 One could easily imagine there being a location in a disc
where the local background pressure gradient is identically zero (because of, say
the global radial distribution of the gas material).  In that case the
effective radial component of gravity is zero and there ought not be any
Rayleigh-Taylor like dynamics present. Under such circumstances, if the density jump across the
interface is non-zero, the consequent activation of self-gravity 
works to stabilize disturbances and its dynamical effect
is to alter the vorticity wave propagation speed only. 
 \par
It is also worth noting that the role of each of the following effects, self-gravity, Rossby and the non-Bousinessq  are to stabilize the flow. This influence is diminished with decreasing wavelength distributions leaving short wavelength unaffected, as can be seen from equation (\ref{total_dispersion_relationship}). The result appears consistent with intuition as the dynamical influence of self-gravity weakens at small scales.
\par
Interestingly, self gravity can also indirectly influence the onset of the Rayleigh-Taylor instability through the basic state (Eq. \ref{base}) and possibly render the flow to be unstable. For example, in the case that the self gravity term  dominates over the inertial terms in the RHS of equation (\ref{base}), one can conclude that these dynamics, which depends strongly on the basic state radial pressure gradient, is mainly due to self gravity contribution.\footnote{\citet{lin2011effect} also report opposite contributions to SG in the perturbed and basic states in their study. However, the system they considered is that of two walls and the interaction could be much more complicated and should be examined in this context.}

\begin{figure}[h!]
\begin{center} 
\fbox{\includegraphics[scale=0.25]{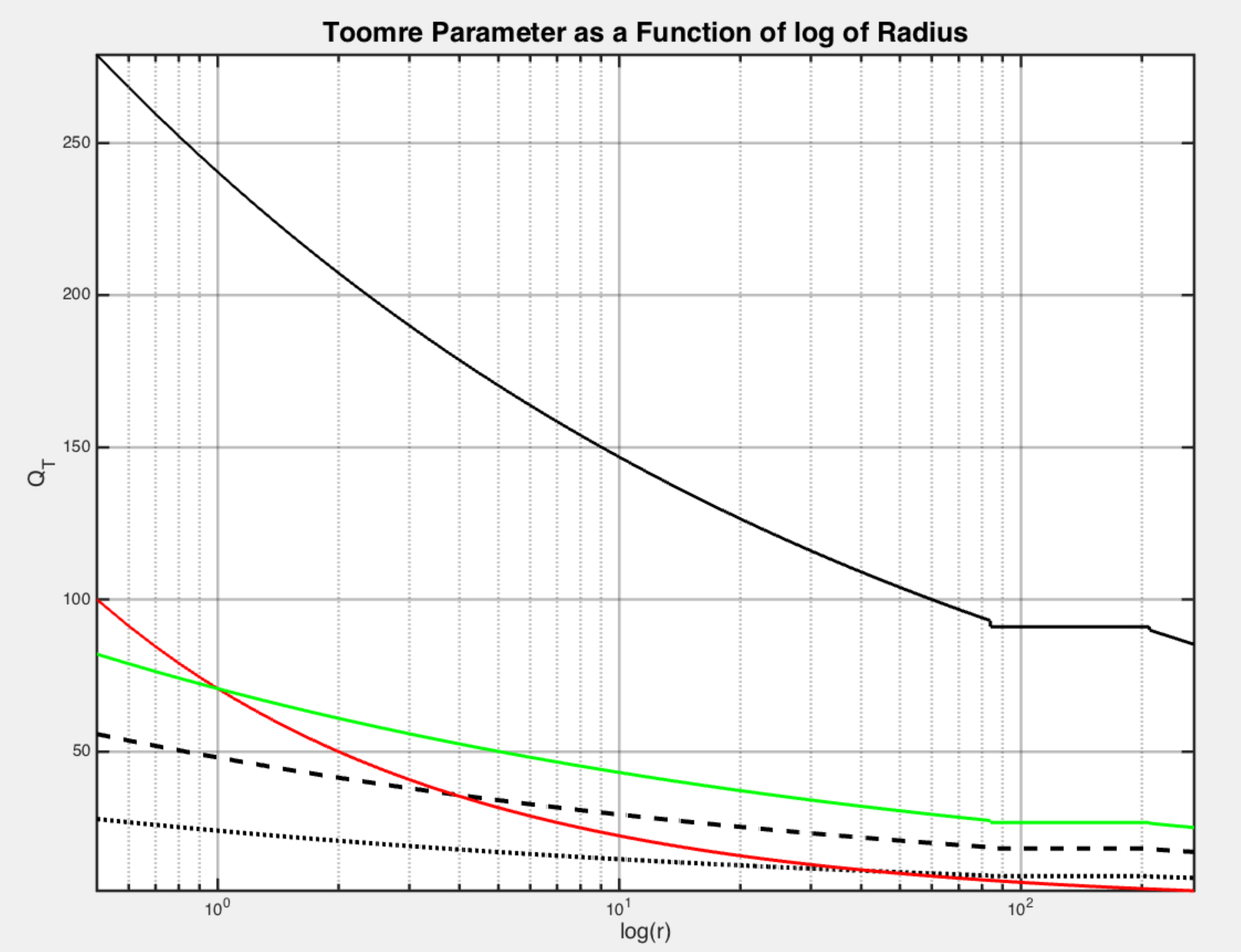}}
\caption{Toomre parameter as a function of the natural logarithm of the radius, for a disk rotating around a $\frac{1}{2}M_{sun}$ star for radii of 0.4-270 [AU].  All lines correspond to radiative hydrostatic equilibrium model of \citet[hereinafter CG97]{chiang1997spectral}, the color indicates what was selected for the calculation of $\frac{h}{r}$, for black $\frac{h}{r}$ is as in CG97, the same for the green line but with $\frac{h}{r}$ adjusted to 0.05 at 1 AU, and in the red line this ratio is constant with constant $\frac{h}{r} =0.05$. Also the full, dashed and dotted lines corresponds to values of $\Sigma$ at 1 AU of $(1,5,10)*10^3 [\frac{gr}{cm^{2}}]$ respectively (where $1*10^3 [\frac{gr}{cm^{2}}]$ correspond to this disk MMSN). It should be noted that the jumps in  side the lines, seen especially in the black lines, is due to the fact the CG97 model isn't continues for $\frac{h}{r}$, and is divided into three regions.}
\label{fig:q}
\end{center}   
\end{figure}

The relevancy of self gravity to the dynamics is usually determined by the value of the Toomre parameter which is defined  as $Q_T=\frac{\Omega_k C_s}{\pi G_{grav} \Sigma}$  for Keplerian disk; where $C_s$ is the local sound speed and $\Sigma$ is the vertically integrated density.  This parameter was originally derived for razor thin disk and it was showed that the disk is stable against gravitational instability for axisymmetric perturbation if $Q_T>1$.  It can be shown that this parameter is related to $\frac{h}{r}\frac{M_*}{M_{disk}}$ \citep{armitage2010astrophysics} (where h is the pressure scale height), which is a more intuitive interpretation of the gravitational instability. It states that the favorable conditions for the onset of the instability are at places in the disk which are flat 
(cold) and heavy.  Figure \ref{fig:q} shows the Toomre parameter as a function of the natural logarithm of the radius. It is easily seen that the expected value of $Q_T$ is highly uncertain and depends on the chosen model (which determines $\frac{h}{r}$) and the ratio of $\frac{M_*}{M_{disc}}$. These quantities are unfortunately unknown both theoretically and observationally. 

Moreover, many studies  have shown that disks can be gravitationally unstable for values of $Q_T$ slightly higher then the classical values. In addition, recent studies have shown that self gravity is important to the overall dynamics even if the value of $Q_T$ is significantly higher then unity. It is known that self gravity has a major effect on the RWI instability in many stages of the dynamics \citep{lin2011effect,LH2013} , and in spiral arms generation \citep{lin2011edge,goldreich1979excitation}. A form of the Q-parameter in this coordinate system  already appears in equation (\ref{total_dispersion_relationship}) 
and is given by

\beqa
{\cal Q}_{T} \equiv \frac{\bar\rho\Omega\sub 0^2}{\pi G_ {grav}\Delta\bar\rho^2},
\nonumber 
\eeqa

It is straightforward to calculate a critical value of ${\cal Q}_{T}$ for the Rayleigh-Taylor instability discussed earlier by equating the two terms in the brackets of the RHS in equation (\ref{sufficient_criterion_for_instability}). It can be seen that ${\cal Q}_{T}$ depends on the relative sizes of the density and vorticity jumps. Nonetheless, one can deduce that SG has an important effect on Rayleigh-Taylor instability even for very high ${\cal Q}_{T}$ values.
We therefore conclude that for a single
interface, instability occurs only if the flow is unstably stratified with respect to the sign of the density jump (Fig. \ref{fig:4a}), and self-gravitational effects should be considered even for Toomre parameter values exceeding one.  
In the absence of instability self-gravity alters the speed of propagation
of the two kernel gravity waves present along the individual interface.  

In a follow-up study we currently examining the role of both self-gravity and non-Boussinesq
effects in the instability mechanism between well separated edge waves (Fig. \ref{fig:4b}), propagating
along locally stably stratified regions, for different Toomre parameter regimes.

\bibliographystyle{gGAF}
\bibliography{gGAFguide}

\end{document}